\documentclass{article}

\usepackage{microtype}
\usepackage{graphicx}
\usepackage{subcaption}
\usepackage{booktabs} 

\usepackage{hyperref}

\usepackage[accepted]{icml2026}

\usepackage{amsmath}
\usepackage{amssymb}
\usepackage{mathtools}
\usepackage{amsthm}
\usepackage{multirow}
\usepackage{longtable}
\usepackage{bm}
\usepackage{supertabular}
\newcommand{\revise}[1]{\textcolor{black}{#1}}

\usepackage{pifont}
\usepackage[capitalize,noabbrev]{cleveref}

\theoremstyle{plain}
\newtheorem{theorem}{Theorem}[section]

\theoremstyle{definition}
\newtheorem{definition}[theorem]{Definition}

\theoremstyle{remark}

\usepackage[disable,textsize=tiny]{todonotes}

\icmltitlerunning{OSM+: Billion-Level OpenStreetMap Dataset for City-wide Experiments}

\begin{document}

\twocolumn[
  \icmltitle{OSM+: Billion-Level OpenStreetMap 
Dataset for City-wide Experiments}



  \icmlsetsymbol{equal}{*}

\begin{icmlauthorlist}
    \icmlauthor{Guanjie Zheng}{sjtu}
    \icmlauthor{Ziyang Su}{sjtu}
    \icmlauthor{Yiheng Wang}{sjtu}
    \icmlauthor{Yuhang Luo}{sjtu}
    \icmlauthor{Hongwei Zhang}{alibaba}
    \icmlauthor{Xuanhe Zhou}{sjtu}
    \icmlauthor{Linghe Kong}{sjtu}
    \icmlauthor{Fan Wu}{sjtu}
    \icmlauthor{Wen Ling}{sjtu}
\end{icmlauthorlist}

  \icmlaffiliation{sjtu}{Shanghai Jiao Tong University}
  \icmlaffiliation{alibaba}{Alibaba Cloud}
  \icmlcorrespondingauthor{Linghe Kong}{linghe.kong@sjtu.edu.cn}
  \icmlcorrespondingauthor{Wen Ling}{wen.ling@sjtu.edu.cn}

  \icmlkeywords{Large Dataset, Traffic Prediction}
  \vskip 0.3in
]



\printAffiliationsAndNotice{}  

\begin{abstract}

Road network data provides rich information about cities, but processing worldwide OpenStreetMap (OSM) data is computationally intensive, and the resulting graphs are often difficult to unify for benchmarking downstream tasks. Existing graph learning benchmarks fail to capture the billion-scale and unique topological properties of real-world road networks, leaving model scalability underexplored. To close this gap, we process OSM data with distributed cloud computing using 5,000 cores and release \textbf{OSM+}, a structured worldwide 1-billion-vertex road network graph dataset designed for high accessibility and usability. OSM+ is open source and globally downloadable, providing an open-box graph structure and an easy spatial query interface; the evaluated release is a fixed snapshot for reproducibility, with a versioned update plan for future releases. We demonstrate the utility of OSM+ through three illustrative use cases: city boundary detection, traffic prediction, and traffic policy control. For traffic prediction, we construct a new 31-city benchmark by processing traffic data and combining it with OSM+, enabling broader spatial coverage and more comprehensive evaluation than commonly used datasets, while scaling from hundreds of road network intersections to thousands. For traffic policy control, we release a new six-city dataset at a much larger scale, introducing challenges for thousand-scale multi-agent coordination. We also provide data processing tools for integrating multimodal spatial-temporal data with OSM+ for geospatial foundation model training, thereby expediting the discovery of compelling scientific insights.

\end{abstract}
\section{Introduction}

Road network has formed the skeleton of cities, as it connects regions within one city and between different cities. For a long time, urban regions and road networks have stretched along each other. Therefore, road networks can essentially reflect the landscape and functional zones in cities, and thus affect human mobility. For instance, it is reported that the human mobility intensity is highly correlated with road network density~\citep{zhu2022understanding}. Hence, investigating road network structures is the basis for urban researches such as urban planning and urban traffic prediction. 

However, obtaining accurate road network data for open public research remains challenging. High-quality road network datasets are collected at high cost by commercial map providers such as Google~\citep{Google}, Bing~\citep{Bing}, Baidu~\citep{Baidu}, and Gaode~\citep{Gaode}. As a result, these services are mainly designed for commercial applications, offering only limited high-level APIs to the public (e.g., POI search, origin-destination route planning~\citep{Google, Baidu}). Such restricted access falls short of the needs of academic researchers and start-ups, who often require fine-grained, low-level access to road network data in order to flexibly conduct computations and rapidly iterate on research ideas or product prototypes.

Open-source map services (e.g., OpenStreetMap~\citep{haklay2008openstreetmap}) built from crowdsourcing mechanisms by worldwide users are promising to resolve this problem. However, due to the massive amount of roadnet data and the complex data format in map object storage, processing the road network data from scratch to obtain the desired format for experiments is always challenging and time-consuming. It might take several trials for human developers (or even with AI) to clean the road network, connect the broken edges, remove the redundant edges, and fix the problems, with each trial costing several hours. Hence, it is highly desirable that \textit{an intermediate format of processed road network data} can support diverse downstream applications so as to speed up the scientific discovery. 

Following this path, some studies~\citep{grinberger2022osm,bartzokas2022utilizing,ding2022time} released tools to clean the OpenStreetMap road network data. However, several issues remain. First, the cleaning of OpenStreetMap contains a complex pipeline, including converting, reducing, transforming and aggregating. This pipeline may take about 10 hours, even only for processing a region with 1,000$km^2$ size, when running on a machine with 32 CPU cores and 128GB memory. Second, the computing of world-wide map data requires far more memory than that of a single machine. The worldwide raw OSM data is roughly 1.1TB before processing, and the finally processed structured OSM data can be a graph with more than one billion vertices. Although segmenting the regions and processing each region separately may make it possible to process the data locally, this requires hand-crafted distributed computing strategies and further aggregating the processed data may introduce extra errors to the data. Because of the two issues mentioned above, researchers have developed different ways to process road network data, which has led to multiple versions being derived from the same raw source. This inconsistency makes it difficult to integrate the data for training multi-modal foundation models. Therefore, it is important to \textit{build a pre-processed, structured version of OpenStreetMap and make it publicly available}.

\begin{figure}[t]
    \centering
    \includegraphics[width=\linewidth]{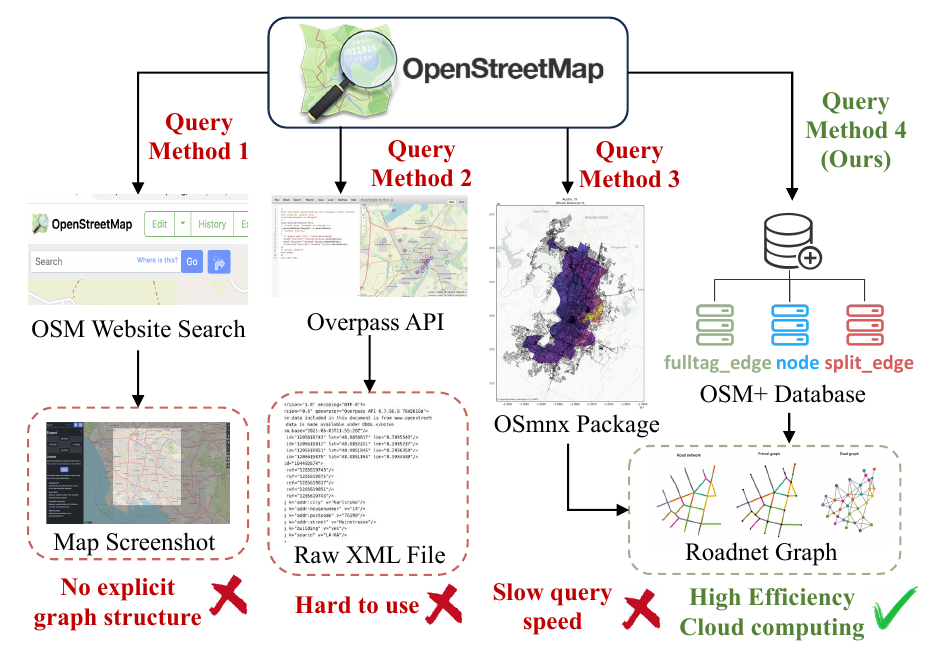}
    \label{fig-intro}
    \caption{Comparison among four ways of querying OpenStreetMap data: website search, the Overpass API, the OSMnx package, and our proposed method (OSM+ dataset). While the first three either lack an explicit graph structure, are difficult to use, or suffer from slow query speed, our OSM+ dataset stored in a cloud database enables efficient, scalable queries over a precomputed graph-structured dataset.}
\end{figure}

In this paper, we introduce OSM+, a dataset, benchmark, and system resource that makes OpenStreetMap usable for large-scale city-wide experiments. OSM+ consists of four released components. \textbf{First}, we release six core OSM+ data tables. Among these six tables, three contain comprehensive non-road information, such as POI nodes, building footprints, and administrative boundaries, while the other three contain the road-network graph used in most experiments. \textbf{Second,} we release an automated preprocessing pipeline that transforms raw Geofabrik/OpenStreetMap mirrors into structured files, making the construction process reproducible. \textbf{Third}, we provide efficient query tools based on the \textit{CityBrain Open Platform} for managing OSM+ and extracting spatial subgraphs at scale. \textbf{Fourth}, we release benchmark code, processed benchmark graphs, experimental configurations, and converters for downstream tasks including city-boundary detection, traffic prediction, and traffic policy control. The release evaluated in this paper uses a static OSM snapshot from March 2026 for reproducibility, and future OSM+ versions will follow a three-month update strategy to track the continuously evolving OSM map.

The contributions of this paper can be summarized as follows:
\begin{itemize}
    \item We introduce OSM+, an open-source, worldwide billion-scale road-network graph dataset with an open-box graph structure and an easy spatial query interface. To our knowledge, OSM+ is the first billion-scale graph resource that enables studying the unique topological properties of real-world road networks and exposes critical scalability limitations of modern GNNs. Detailed instructions for accessing the OSM+ dataset are shown in Appendix \ref{data_access}.
    \item We provide a set of cloud-computing-based APIs to enable efficient billion-scale graph query and processing, as well as an automated pipeline for transforming raw Geofabrik/OpenStreetMap files into structured node, split-edge, and full-tag tables. The six OSM+ tables are available at \url{https://huggingface.co/datasets/suziyang/OSM_plus}, while the related code, preprocessing scripts, query tools, tutorials, benchmark code, and converters are available at \url{https://github.com/SJTU-CILAB/OSM-dataset}.
    \item We provide four example application scenarios based on the OSM+ dataset: query operations, city-boundary detection, traffic prediction, and traffic policy control. The released converters show how OSM+ can serve as a road-network backbone for integrating multi-modal spatial-temporal data and for constructing reproducible downstream tasks.
    \item For the traffic-prediction task, we construct a new 31-city offline research benchmark by aligning public historical traffic-flow observations from UTD19 with OSM+ road networks. For traffic policy control, we release six city-scale simulation scenarios with up to 18,948 intersections. These resources raise new challenges for both predictive accuracy and scalability, while source-specific limitations of the third-party upstream datasets are explicitly documented. 
\end{itemize}

\section{OSM+: Structured Road Network Dataset with Computing Support}

\begin{figure*}[htbp]
    \centering
    \includegraphics[width=0.98\textwidth]{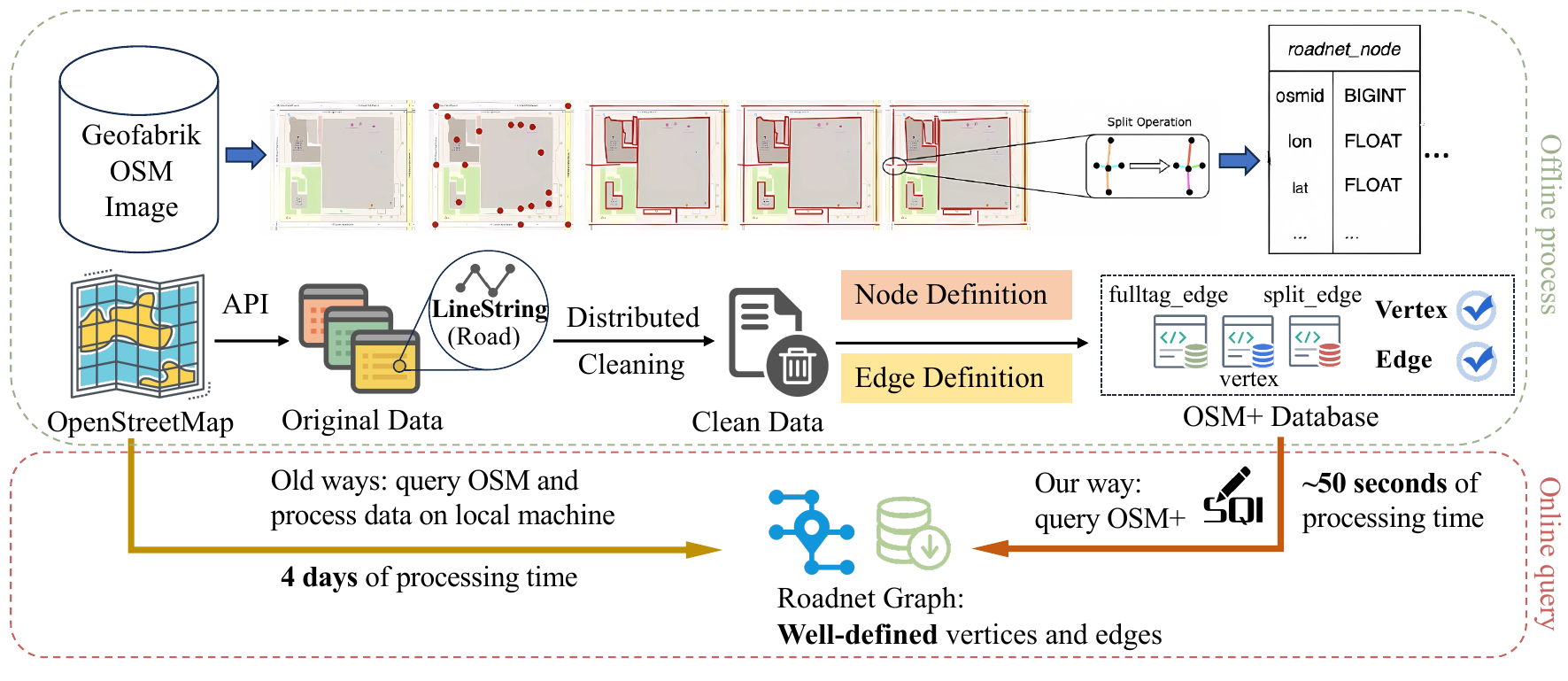}
    \caption{OSM+ database provides worldwide edges (Table split\_edge) and points (Table vertex). Only basic SQL is needed to facilitate efficient construction of urban graph structures at global scale.}
    \label{osm-plus-overview}
\end{figure*}

To address the challenges of raw OSM data usability, we present OSM+, a scalable system that transforms raw map data into structured, analysis-ready road networks. As shown in Figure~\ref{osm-plus-overview}, in this section, we separately introduce the key components of OSM+, including graph-based network modeling, efficient spatial query operations, and adaptive data converters.

\subsection{Graph-Based Road Network Modeling}
A lot of downstream machine learning tasks in the transportation domain  (e.g., traffic prediction) rely on road network graph structures. Thus, we generalize the road network as a graph with billions of vertices and edges under the following definition. Our OSM+ dataset is a graph-structured database based on MaxCompute, formerly called ODPS. MaxCompute is Alibaba Cloud’s fully managed big data computing and data warehouse platform. It can store and process terabytes to petabytes of data, offering SQL and other distributed computing models to analyze massive datasets efficiently while reducing costs and ensuring strong data security. Here is the definition of our roadnet graph model:

\begin{definition}[Graph-based Road Network Model]
Given a city-scale road network with intersections and road segments, we can define a graph structure with the vertices and edges defined as follows.
\item{\it (1) \textbf{Vertex}: Each vertex represents a road intersection in the OpenStreetMap road network or POI point.}
\item{\it (2) \textbf{Edge}: Each edge represents a road segment in OpenStreetMap with a starting intersection and ending intersection. Since each road segment may have multiple parts segmented by minor intersections or direction turning points in geometry, each edge may contain multiple line segments.}
\end{definition}

This graph model can facilitate quick queries and seamless table joining. For example, a single SQL query can be issued to retrieve all road segments within a specified bounding box and join them with a POI table to compute features such as the number and types of nearby facilities (e.g., schools, gas stations, or shopping malls) for each road segment.

\subsection{Basic Statistics of OSM+}
\textbf{Statistics.} In the original OpenStreetMap, a road is represented as a linestring. We subdivide each linestring into several split edges, each with exactly one start point and one end point. The split edges that constitute the road network are referred to as “split edges on roadnet”. It is straightforward to observe that the worldwide road network forms a massive graph with billions of vertices and edges. Moreover, the spatial density of vertices can serve as a proxy for population distribution and levels of urbanization. The overall statistics are shown in Table~\ref{basic statistics}.

\begin{table}[h]
\centering
\renewcommand{\arraystretch}{1.2}
\setlength{\tabcolsep}{3pt}
\caption{Some basic statistics of OSM+ database.} 
\label{basic statistics}
\resizebox{\linewidth}{!}{
    \begin{tabular}{c|c|c|c|c}
    \toprule
    \multirow{2}{*}{\textbf{Subset}} 
        & \multicolumn{3}{c|}{\textbf{Counts}} 
        & \textbf{Total} \\
    \cline{2-4}
        & \textbf{Vertices} 
        & \textbf{Split Edges} 
        & \textbf{Linestrings} 
        & \textbf{Length (km)}\\ 
    \hline
    \textbf{Roads and POI} 
        & 9,864,616,444 
        & 10,656,542,261
        & 1,125,602,655
        & \multirow{2}{*}{84,662,999} \\
    \cline{1-4}
    \textbf{Roadnet Only} 
        & 2,180,447,343 
        & 1,964,857,157 
        & 197,775,476
        &  \\
    \bottomrule
    \end{tabular}
    }
\end{table}

\begin{figure}[ht]
    \centering
    \includegraphics[width=\linewidth]{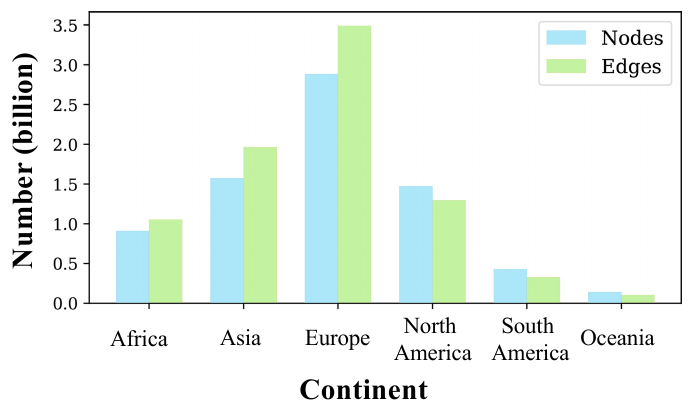}
    \caption{Basic continent-level statistics of the OSM+ database, showing the number of road-network nodes and edges across different continents.}
    \label{histogram-continent}
\end{figure}

\textbf{Spatial distribution \& categories.} The spatial distribution of road data is highly imbalanced. Figure~\ref{histogram-continent} reports the numbers of roadnet vertices and edges for each continent. Europe exhibits the largest number of vertices and edges, followed by Asia, while Oceania has the smallest. Furthermore, we compute the total length of the road network in five representative countries or regions, and we collect the corresponding 2019 road-length statistics from the International Road Federation. These statistics are presented in Table~\ref{five countries}. By comparing these data, we observe that OSM road coverage in the United States, the Russian Federation, and Canada is more complete, whereas coverage in China and India remains relatively incomplete. This comparison should be interpreted as a coverage audit rather than a claim that OSM+ is uniformly complete across all regions. The distribution of road categories is shown in Figure~\ref{histogram-continent}. We observe that the number of roads increases approximately exponentially as the road hierarchy level decreases, which provides a compact view of the hierarchy-tag distribution inherited from OSM.

\begin{table}[ht]
    \centering
    \renewcommand{\arraystretch}{1.2}
    \caption{Total length of road (km) in different countries from OSM+ and IRF data sources.}
    \label{five countries}
    \resizebox{\linewidth}{!}{
        \begin{tabular}{c|c|c|c|c|c}
        \toprule
        \textbf{Data Source} & \textbf{China} & \textbf{America} & \textbf{Russia} & \textbf{India} & \textbf{Canada}\\
        \hline 
        \textbf{OSM+} & 3,805,919 & 13,731,271 & 9,956,729 & 2,878,553 & 1,592,642 \\
        \hline
        \textbf{IRF Statistics}& 5,012,496 & 6,638,329 & 1,542,196 & 6,371,847 & 1,126,600  \\
        \bottomrule
        \end{tabular}
    }
\end{table}

\begin{figure}[ht]
    \centering
    \includegraphics[width=\linewidth]{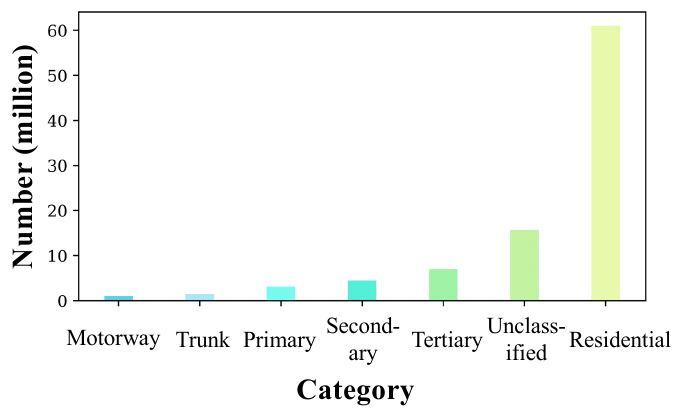}
    \caption{Category-level statistics of the OSM+ database, illustrating the distribution of road segments across major road types, including motorway, trunk, primary, secondary, tertiary, unclassified, and residential roads.}
    \label{histogram-category}
\end{figure}

\section{Basic Usage of OSM+}
\subsection{Efficient Spatial Computing Support}

\begin{figure}[ht]
    \centering
    \includegraphics[width=\linewidth]{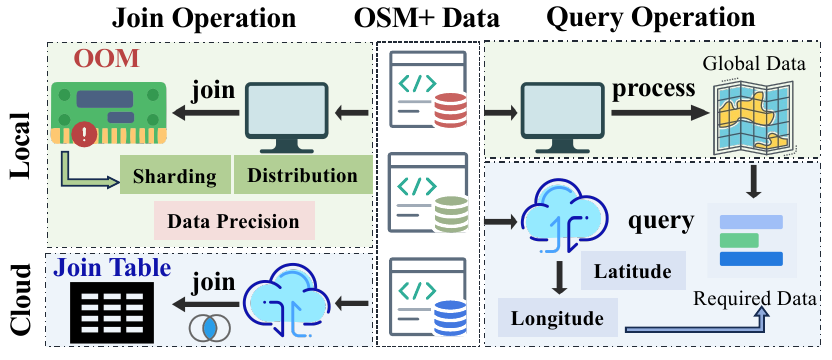}
    \caption{Comparison between conventional local OSM processing and the OSM+-based cloud query workflow.}
    \label{query-workflow-comparison}
\end{figure}

Given that OSM+ encodes a global road network with billions of vertices and edges, executing interactive queries on a single commodity server is impractical. We therefore deploy OSM+ on cloud-computing resources and design parallel query mechanisms to enable efficient large-scale access to the data. Figure~\ref{query-workflow-comparison} illustrates the difference between the conventional local processing workflow and the OSM+-based cloud query workflow. In the local workflow, users typically need to download and process large raw OSM tables, perform expensive join operations, and manually handle sharding, distribution, and data-precision issues. When the target region is large or the join involves global-scale tables, this process can easily exceed the memory capacity of a local machine and lead to out-of-memory errors. Moreover, even if only a specific regional map is needed, the local workflow still requires cumbersome preprocessing over a much larger data scope before the required subset can be extracted.

In contrast, OSM+ organizes the processed global road-network data in the cloud and exposes efficient query interfaces. Users only need to specify the query condition, such as longitude-latitude coordinates or a target spatial region, and the system automatically locates the relevant shards, executes the query in parallel, and returns the required data. This design avoids repeated local preprocessing, reduces the memory burden on users, and makes large-scale spatial access to OSM+ substantially easier and more efficient. Here, we use several basic query operations as examples to demonstrate how OSM+ supports practical access to the global road-network database.

\textbf{Point Query Operation.} Point query is one of the fundamental spatial operations supported by OSM+. Given a query point and a radius, the goal is to select all vertices in the \texttt{osm\_node} table that fall within the specified radius. We apply window-check optimization that partitions the geographic space into a uniform grid at a specified spatial resolution. During query processing, we restrict the search to vertices located in the grid cell containing the query point and its neighboring cells, thereby pruning a large fraction of irrelevant vertices. As shown in Table~\ref{window check}, we conduct 1,000-point queries with and without the window-check optimization on a commodity cloud-computing cluster, and the results indicate that window check substantially improves point-query efficiency.

\textbf{Nearest-neighbor Query Operation.} Given a query point and a latitude-longitude bounding box, the goal is to find the nearest vertex (nearest-neighbor query) or the $k$ nearest vertices ($k$-nearest-neighbor query) within this region. To reduce query time, we employ $k$-d tree, a binary space-partitioning tree that recursively divides a $k$-dimensional space. As shown in Table~\ref{window check}, the $k$-d tree implementation substantially reduces the running time required for these queries compared with a linear scan baseline.

\begin{table}[h]
    \centering
    \renewcommand{\arraystretch}{1.2}
    \caption{Query efficiency comparison between with or without “window check” and “$k$-d tree” optimization technique.}
    \label{window check}
    \resizebox{\linewidth}{!}{
        \begin{tabular}{c|c|c|c|c}
        \toprule
        \multirow{2}{*}{\textbf{Method}} & \multicolumn{2}{c|}{\textbf{Point Query}} & \multicolumn{2}{c}{\textbf{Nearest Pair Query}} \\
        \cline{2-5}
        & Runtime(s) & Core $\times$ min & Runtime(s) & Core $\times$ min \\
        \hline

        \textbf{W/O Optimization} & 310 & 2.06 & 41.14 & 0.05 \\
        \hline
        \textbf{With Optimization} & \textbf{79} & \textbf{1.33} & \textbf{1.49} & \textbf{0.01} \\
        \bottomrule
        \end{tabular}
        }
\end{table}

\begin{table}[htb]
    \centering
    \renewcommand{\arraystretch}{1.2}
\caption{Efficiency of global KDE estimation on different computing clusters.}
\label{odps_efficiency}
\resizebox{\linewidth}{!}{
\begin{tabular}{c|c|c|c|c|c|c}
\toprule

\multirow{3}{*}{\textbf{Platform}} & \multicolumn{6}{c}{\textbf{Sample Rate}}\\ \cline{2-7}
& \multicolumn{2}{c|}{\textbf{1/1,000}} & \multicolumn{2}{c|}{\textbf{1/10,000}} & \multicolumn{2}{c}{\textbf{1/100,000}}\\
\cline{2-7}
& Runtime(s) & Core $\times$ min & Runtime(s) & Core $\times$ min & Runtime(s) & Core $\times$ min \\
\hline
\textbf{ECS} & OOM & OOM & 841.62 & 14.02 & 33.06 & 0.22 \\
\hline
\textbf{Spark} & 29,014.66 & 17,408.49 & 197.02 & 118.21 & 7.21 & 4.32 \\
\hline
\textbf{ODPS} & \textbf{72.13} & \textbf{1.02} & \textbf{34.26} & \textbf{0.30} & \textbf{6.04} & \textbf{0.10} \\
\bottomrule
\end{tabular}
}
\end{table}

Using these techniques, we build an example comprehensive calculation task to conduct the KDE kernel density estimate of each intersection vertex on the global road network data, illustrating why it is necessary to run experiments on cloud computing. We use three different sampling rates to sample the original global road network data and compare the runtime and memory cost on different platforms. The experimental results are shown in Table~\ref{odps_efficiency}. It is observed that by utilizing the ODPS computing engine, we could employ an optimized query algorithm which significantly outperforms that of other computing platforms in terms of both runtime and memory utilization. 

\subsection{Use Case: City Boundary Detection}

\begin{figure*}[ht]
    \centering
    \includegraphics[width=0.96\linewidth]{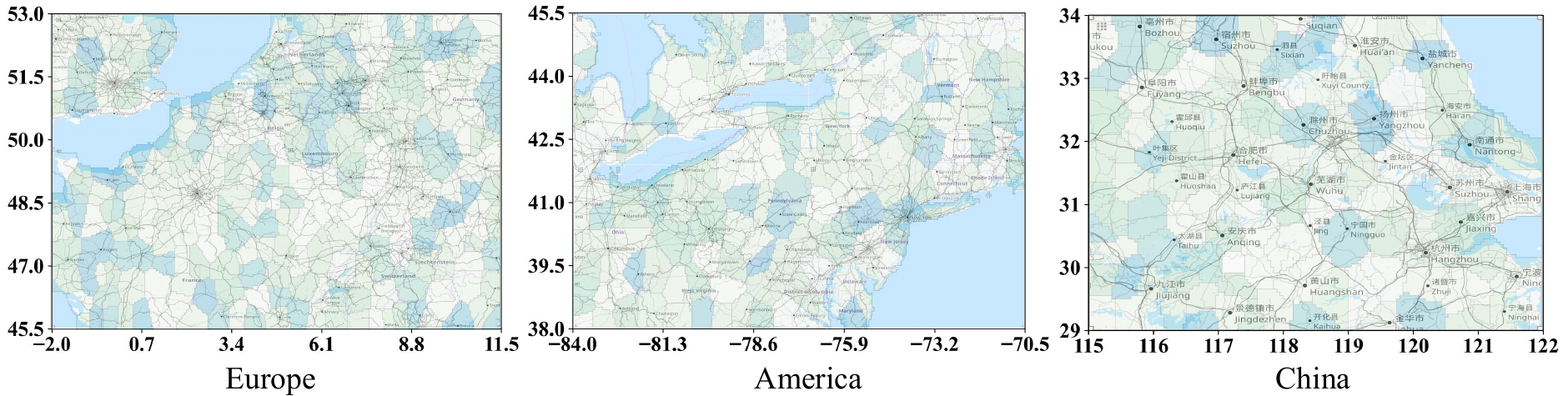}
    \caption{The city boundary map obtained from the clustering results of central Europe, the east coast of the United States and the Yangtze River Delta in China is overlaid with the road maps of the three regions on OpenStreetMap, and latitude and longitude are added to distinguish them.}
    \label{fig:part}
\end{figure*}

\begin{figure*}[ht]
    \centering
    \includegraphics[width=\linewidth]{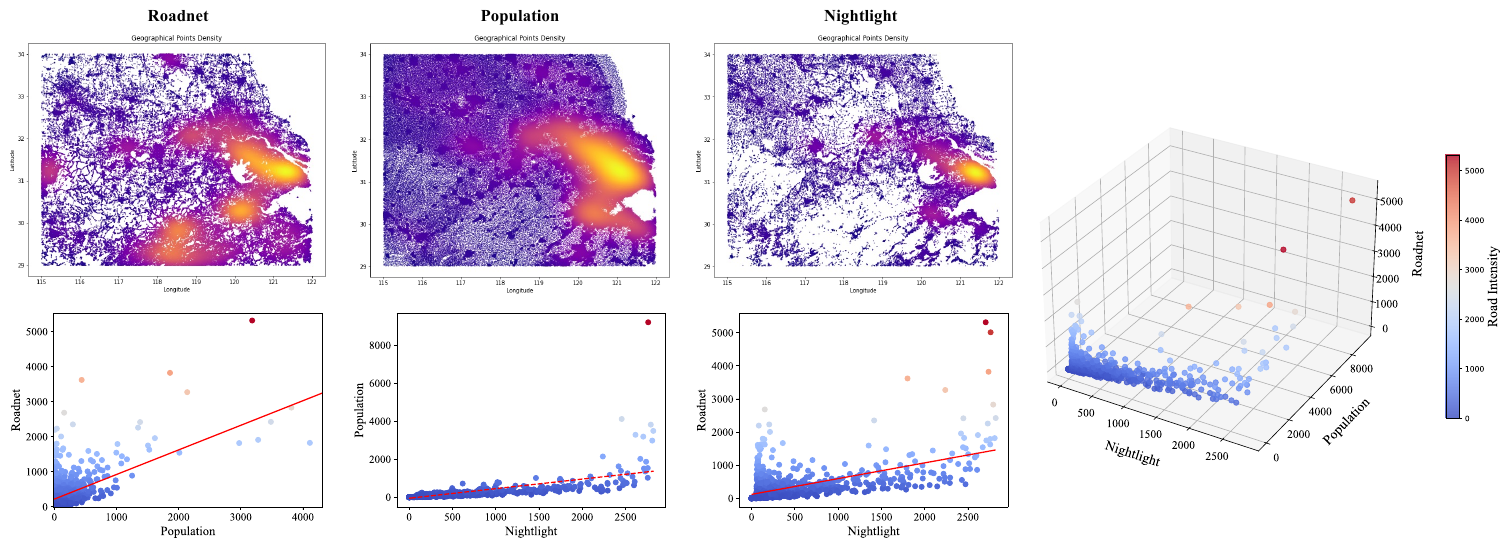}
    \caption{Comparing the road, population and nightlight data map in the same area, the road density is positively correlated with the population and night lighting data density.}
    \label{fig:otherdata}
\end{figure*}

Accurately delineating city boundaries is important for urban planning, public resource allocation, and location-based services. However, administrative boundaries do not always reflect the functional extent of cities, especially in rapidly urbanizing regions where economic activities and commuting flows may extend beyond official borders. Since urban areas are typically characterized by dense and interconnected road networks, road-network density provides a useful signal for identifying functional city boundaries.

In this use case, we show how OSM+ can be used to derive functional city boundaries from global-scale road-network data. Specifically, we cluster regions according to road-network structure and density, and compare the resulting clustering-based boundaries with official administrative boundaries from the \textbf{Database of Global Administrative Areas}~\citep{GADM}. By overlaying the two types of boundaries, we identify both consistent regions and mismatched areas. Large discrepancies indicate that the effective urban footprint has expanded beyond administrative limits or that one administrative unit contains multiple relatively separated urbanized areas.

Figure~\ref{fig:part} presents representative results in central Europe, the east coast of the United States, and the Yangtze River Delta in China. Colored polygons denote the clustering-based city regions, while boundary lines indicate administrative divisions. In many cases, such as Shanghai and Jiaxing, the clustering results are broadly consistent with administrative boundaries. In other cases, an administrative region may contain multiple clusters, suggesting several disconnected urban centers within the same jurisdiction. Mismatches are often observed near administrative borders where dense road networks continue across official boundaries, reflecting cross-border urban expansion or strong regional connectivity.

To further validate the road-density-based interpretation, we compare road networks with population and nightlight data in Figure~\ref{fig:otherdata}. The results show that areas with denser roads generally coincide with higher population density and stronger night-time light intensity, suggesting that road-network density is a reasonable proxy for urban activity. In addition, population maps can be thresholded by high red-channel values in RGB space to approximate city-center regions, while nightlight maps can be filtered by brightness to extract highly active urban cores. These comparisons indicate that OSM+ can support scalable and interpretable city boundary detection by providing a unified road-network foundation.
\section{OSM+ Assisted Tasks}

\begin{figure}[ht]
    \centering    
    \includegraphics[width=\linewidth]{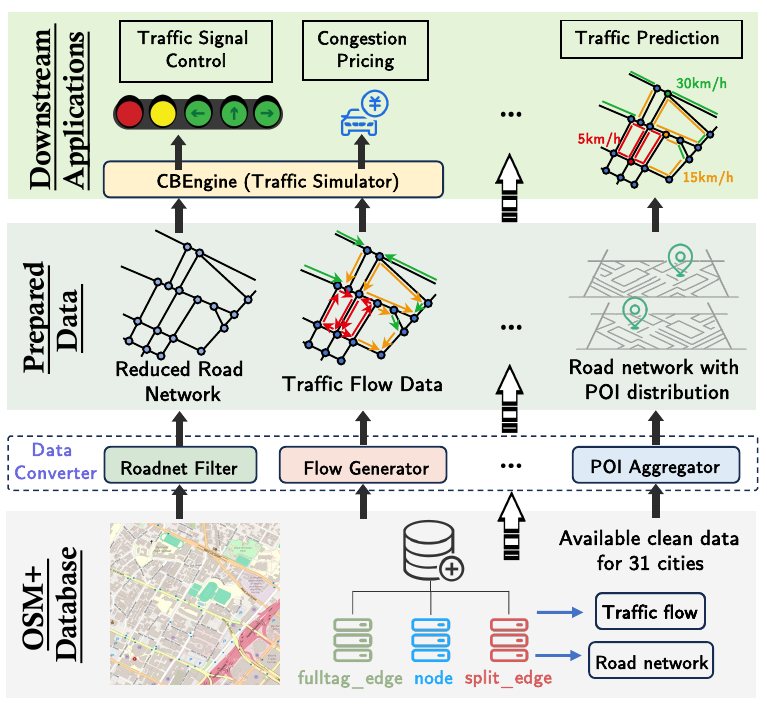}
    \caption{The framework of the downstream task supporting OSM+ database.}
    \label{fig:framework}
\end{figure}

In this section, we introduce two typical applications of OSM+ combined with other datasets, which are \textbf{traffic prediction} and \textbf{traffic policy control}. As summarized in Figure~\ref{fig:framework}, OSM+ provides a unified data foundation for downstream urban and transportation applications. Starting from the OSM+ database, the pipeline first uses task-specific data converters to transform raw road-network records and auxiliary urban information into application-ready inputs. The prepared data can then be directly consumed by downstream systems such as CBEngine for traffic simulation, supporting tasks including traffic signal control and congestion pricing, or used as structured spatial inputs for traffic prediction. In this way, OSM+ bridges large-scale road-network data and practical downstream tasks through a standardized conversion-and-application workflow.

\begin{figure*}[t]
    \centering
    \setlength{\abovecaptionskip}{-0.1cm}
    \includegraphics[width=\textwidth]{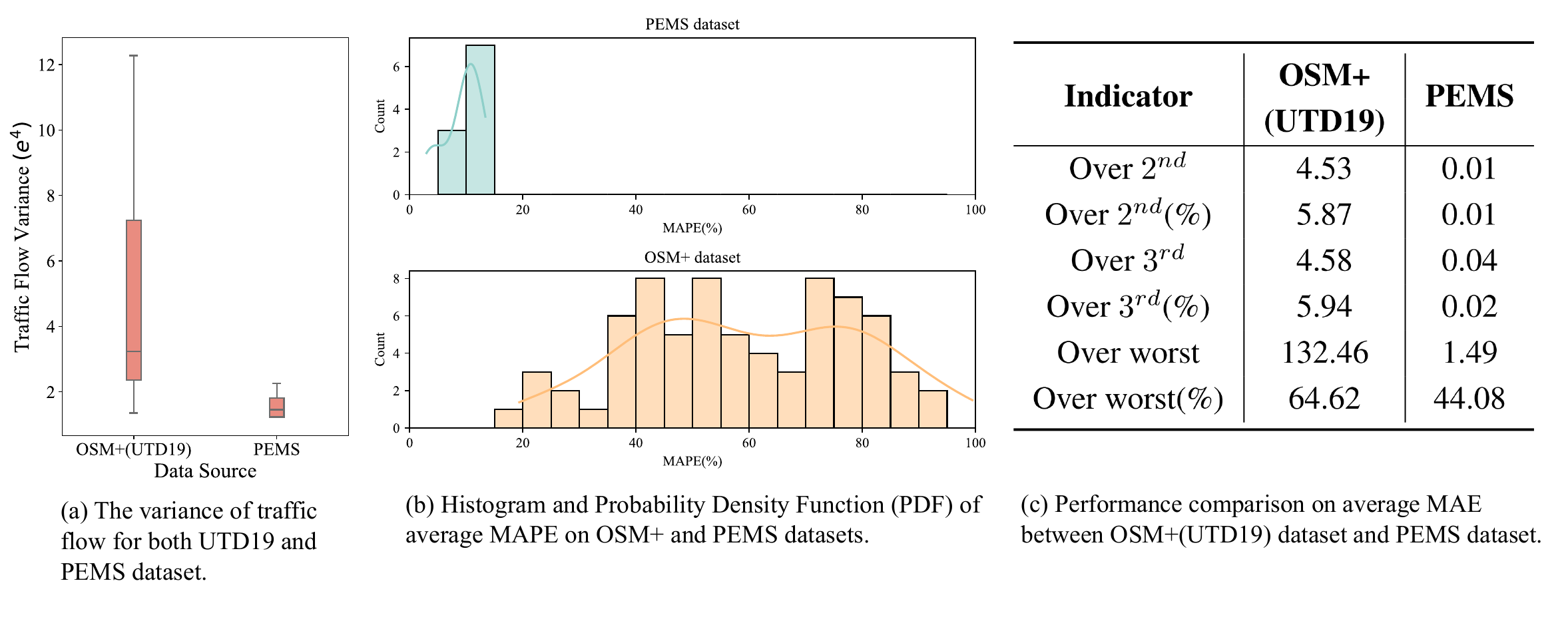}
    \caption{Comparison between the newly constructed OSM+(UTD19) dataset and PEMS dataset in three aspects.}
    \label{fig:compare}
\end{figure*}
\begin{table*}[ht]

\caption{Basic statistics of traffic signal control experiments.}
    \centering
    \renewcommand{\arraystretch}{1.2}
    \setlength{\tabcolsep}{4pt}
\label{traffic signal control}
\resizebox{0.95\linewidth}{!}{
\begin{tabular}{c| c | c | c | c | c | c | c | c}
\toprule
 \multirow{2}{*}{\textbf{Indicator}}& \multicolumn{2}{c|}{\textbf{Former datasets}} & \multicolumn{6}{c}{\textbf{OSM+ dataset}} \\
\cline{2-3} \cline{4-9}
~ & Hangzhou & Manhattan & New York & Los Angeles & Beijing & Shanghai & London & Paris\\
\hline
Intersections & 16 & 2,510 & 5,971 & 6,663 & 18,948 & 14,750 & 5,895 & 1,721 \\
\hline
Vehicles & 2,983 & 48,079 & 90,059 & 112,291 & 130,851 & 85,480 & 107,105 & 101,929 \\
\bottomrule
\end{tabular}
}
\end{table*}

\subsection{City-scale Traffic Prediction}

Traffic prediction has been extensively studied due to its importance for intelligent transportation systems. A large body of work evaluates models on standard benchmarks such as METR-LA and the various PEMS datasets~\cite{li2017diffusion,jiang2022graph}. However, recent studies have noted that on these static benchmarks, many newly proposed methods achieve only marginal improvements over strong baselines, raising concerns about benchmark saturation and overfitting~\cite{oakley2024foresight}.

Thus, we propose \textbf{31 new city-level datasets} associated with traffic flow data, to provide a more comprehensive benchmark for this problem. In addition to the many more cities compared with previous datasets, the newly proposed datasets are different from previously used datasets for several reasons, as shown in Figure~\ref{fig:compare}.
\begin{itemize}
    \item \textbf{Dynamic in-city Scenario:} Unlike previous datasets, which may have focused on more uniform highway conditions, these datasets capture a broader spectrum of in-city scenarios. This introduces greater variability in the data, reflecting the diverse and dynamic nature of urban environments. Such variability is crucial for developing models that can effectively handle the complexities and unpredictability inherent in city traffic patterns and infrastructure.
    \item \textbf{Sparsity Challenge:} The number of sensors is relatively low compared to the road intersections or segments. Modeling sparsity is critical as it mirrors real-world conditions where data points can be irregular or missing. Addressing sparsity effectively can significantly improve the accuracy and reliability of the model, ensuring it performs well even in less-than-ideal data conditions. This aspect of the dataset pushes the boundaries of current modeling techniques, encouraging the development of more sophisticated and resilient algorithms.
    \item \revise{\textbf{Large-scale Coverage:} Our OSM+ dataset also operates at a substantially larger spatial scale than widely used traffic benchmarks. While previous widely used datasets such as PEMS and METR-LA typically contain only a few hundred sensor nodes, our 31 city-level datasets are defined over road networks with a much larger number of intersections and segments; in particular, the largest city graph contains approximately 30{,}000 nodes. This large-scale setting provides a more realistic and challenging test bed for evaluating not only predictive accuracy but also the scalability of traffic models in city-wide deployment scenarios.}
\end{itemize}

\begin{table*}[t]
    \centering
    \renewcommand{\arraystretch}{1.09}
    \caption{Traffic-prediction results on 31 OSM+(UTD19) city datasets with seven baseline methods. Values are averaged over horizons 3, 6, and 12. MAE and MAPE(\%) are reported; the best value in each row is bolded. OOM indicates that the run exceeded the 24GB GPU memory limit under our unified setting.}
    \label{tab:traffic prediction}
    \resizebox{\linewidth}{!}{
    \begin{tabular}{c|c|c|c|c|c|c|c|c|c|c|c|c|c|c}
    \toprule
        \multirow{2}{*}{  \textbf{City}  } & \multicolumn{2}{c|}{\textbf{AGCRN}} & \multicolumn{2}{c|}{\textbf{Crossformer}} & \multicolumn{2}{c|}{\textbf{DCRNN}} & \multicolumn{2}{c|}{\textbf{DLinear}} & \multicolumn{2}{c|}{\textbf{FEDformer}} & \multicolumn{2}{c|}{\textbf{GWNet}} & \multicolumn{2}{c}{\textbf{MTGNN}} \\ 
    \cline{2-15}
        & MAE & MAPE & MAE & MAPE & MAE & MAPE & MAE & MAPE & MAE & MAPE & MAE & MAPE & MAE & MAPE \\ 
    \hline
        AGB & 47.92  & 44.03  & \textbf{44.40}  & \textbf{35.34}  & OOM  & OOM  & 47.80  & 37.84  & 51.94  & 45.19  & 47.06  & 37.14  & 46.97  & 40.47  \\ 
        BSL & 64.51  & 55.38  & 62.82  & 65.68  & 119.15  & 184.45  & 61.50  & \textbf{51.74}  & \textbf{59.37}  & 61.78  & 81.07  & 106.88  & 78.41  & 93.81  \\ 
        BRN & 51.00  & 231.93 & \textbf{49.84}  & \textbf{201.84} & OOM  & OOM  & 52.18  & 253.27  & 55.58  & 248.00  & 50.59  & 319.09  & 70.90  & 405.07  \\ 
        BHX & 112.08  & 70.94  & \textbf{84.13}  & \textbf{48.08}  & 303.46  & 195.15  & 111.22  & 66.91  & 119.52  & 65.15  & 107.09  & 66.83  & 91.68  & 49.44  \\ 
        BOL & \textbf{31.27}  & 21.07  & 32.74  & \textbf{20.98}  & 38.00  & 26.82  & 37.27  & 34.12  & 37.87  & 29.83  & 35.03  & 25.67  & 32.31  & 21.12  \\ 
        BOD & 71.65  & 39.69  & 67.13  & \textbf{36.19}  & 232.07  & 276.51  & \textbf{67.13}  & 44.54  & 70.14  & 46.29  & 74.18  & 57.18  & 89.14  & 56.70  \\ 
        BRE & \textbf{56.31}  & 36.52  & 58.08  & \textbf{34.22}  & OOM & OOM & 63.27  & 42.47  & 61.42  & 41.98  & 57.01  & 36.98  & 56.69  & 35.50  \\ 
        KN & OOM & OOM & 44.78  & \textbf{48.85}  & 117.40  & 292.38  & \textbf{38.69}  & 61.19  & 40.98  & 67.71  & 43.85  & 55.18  & 47.89  & 75.01  \\ 
        DA & 57.22  & 51.76  & \textbf{53.28}  & 50.75  & OOM & OOM & 54.76  & 53.99  & 57.41  & 61.16  & 54.69  & 51.74  & 57.20  & \textbf{50.30}  \\ 
        ESS & 41.95  & \textbf{34.65}  & 40.47  & 41.68  & 174.64  & 294.77  & 50.35  & 43.85  & 44.70  & 41.95  & 38.99  & 34.87  & \textbf{38.41}  & 36.46  \\ 
        FRA & 163.16  & 54.95  & 145.88  & 47.61  & 179.03  & 51.37  & \textbf{99.62}  & \textbf{30.19}  & 107.78  & 31.76  & 190.85  & 62.23  & 284.10  & 92.89  \\ 
        GRZ & 61.15  & 113.62  & \textbf{52.78}  & \textbf{66.74}  & 183.88  & 464.62  & 60.83  & 72.86  & 56.16  & 73.15  & 58.03  & 68.32  & 56.60  & 74.38  \\ 
        GRQ & 69.64  & 35.57  & 68.02  & \textbf{32.53}  & 158.26  & 114.53  & \textbf{66.03} & 37.54  & 79.09  & 42.20  & 67.99  & 34.95  & 74.99  & 39.00  \\ 
        HAM & 46.50  & 44.89  & 44.49  & 44.87  & 97.85  & 108.12  & 46.69  & 49.81  & 47.85  & 50.69  & \textbf{44.25}  & \textbf{43.83}  & 45.02  & 44.18  \\ 
        INN & 72.80  & 31.56  & 69.28  & 37.40  & 342.05  & 314.50  & 89.95  & 39.55  & 74.44  & 32.32  & \textbf{67.03}  & \textbf{28.53}  & OOM & OOM \\ 
        KS & 81.26  & 106.06  & 86.38  & 118.43  & 233.90 & 427.98  & 75.29  & 107.43  & 89.83  & 127.22  & \textbf{71.23}  & \textbf{94.88}  & 191.45  & 316.68  \\ 
        MAN & 106.16  & 42.54  & 97.48  & 41.10  & 336.42  & 280.35  & 101.38  & 46.21  & 110.81  & 52.15  & \textbf{95.91}  & \textbf{38.95}  & 97.30  & 40.74  \\ 
        MEL & 50.24  & 45.88  & \textbf{45.36}  & 42.73  & OOM & OOM & 63.72  & 66.55  & 53.25  & 56.25  & 51.91  & \textbf{36.10}  & 45.48  & 40.26  \\ 
        RTM & \textbf{52.48}  & \textbf{40.29}  & 53.83  & 50.52  & 179.76 & 347.68 & 68.83  & 53.19  & 68.43  & 65.17  & 67.03  & 50.91  & 57.34  & 41.07  \\ 
        SDR & 103.63  & 59.74  & 102.25  & 65.34  & 262.60 & 271.61  & 97.97  & 54.70  & 125.51  & 95.71  & \textbf{89.36}  & 47.38  & 97.54  & \textbf{44.61}  \\ 
        SP & 49.08  & 39.57  & \textbf{47.93}  & 37.74  & 119.56  & 119.22  & 52.95  & 45.39  & 53.42  & 44.88  & 48.34  & 38.34  & 48.05  & \textbf{37.48}  \\ 
        SXB & 78.34  & 39.40  & 76.17  & 38.72  & 261.11 & 223.11 & 85.62  & 46.72  & 84.71  & 46.10  & 76.86  & 39.46  & \textbf{76.01}  & \textbf{37.36}  \\ 
        STR & 58.93  & 20.37  & 56.60  & 19.52  & 68.19 & 23.30 & 65.80  & 24.52  & 68.38  & 23.48  & \textbf{55.80}  & \textbf{19.05}  & OOM & OOM \\ 
        TPE & 136.50  & 48.04  & 134.51  & 48.18  & 502.95 & 274.25  & 142.61  & 46.21  & 149.12  & 53.31  & \textbf{129.13}  & \textbf{40.14}  & 130.36  & 41.42  \\ 
        TO & 89.48  & 57.66  & \textbf{81.70}  & \textbf{44.44}  & 314.62 & 390.29 & 85.13  & 48.01  & 87.85  & 56.18  & 102.69  & 60.64  & 104.28  & 68.82  \\ 
        YTO & 51.73  & 39.35  & \textbf{51.54}  & 40.18  & 161.46  & 145.72  & 90.53  & 71.76  & 62.92  & 59.10  & 58.04  & 38.73  & 52.24  & \textbf{37.42}  \\ 
        TLS & 257.82  & 751.49  & 255.29  & 756.09  & 268.70 & 847.32 & 263.95  & 870.21  & 296.55  & 836.03  & \textbf{255.26}  & 751.62  & 258.62  & \textbf{730.09}  \\ 
        UTC & OOM & OOM & 50.35  & 62.80  & OOM & OOM & 50.78  & 54.42  & 66.80  & 88.25  & 74.98  & 88.33  & \textbf{39.92 } & \textbf{36.74}  \\ 
        VNO & 88.95  & 54.81  & 84.09  & 49.34  & OOM & OOM & 76.03  & 43.69  & 88.84  & 49.53  & \textbf{73.80}  & \textbf{39.27}  & 96.47  & 64.87  \\ 
        WOB & 54.48  & 41.34  & \textbf{52.21}  & \textbf{39.71}  & 59.39  & 47.61  & 62.24  & 50.94  & 57.60  & 50.15  & 54.32  & 42.30  & 53.24  & 40.17  \\ 
        ZRH & OOM & OOM & 54.73  & 36.93  & OOM & OOM & 60.36  & 43.84  & 60.12  & 43.74  & 66.51  & 53.31  & \textbf{53.52}  & \textbf{35.16}  \\ 
        \hline
        \textbf{\# Win} & 3 & 2 & 10 & 10 & 0 & 0 & 4 & 2 & 1 & 0 & 9 & 8 & 4 & 8 \\
    \bottomrule
    \end{tabular}
    }
\end{table*}

To evaluate the performance, we test seven frequently-cited baseline methods on these 31 cities. All models are evaluated on three different prediction horizons (3, 6, and 12) following the widely used setting. The average results over prediction horizons are shown in Table~\ref{tab:traffic prediction}, and the full per-horizon results with standard deviations are shown in Appendix~\ref{extended results}. Details of all baseline methods are shown in Appendix~\ref{traffic prediction details}.

The results confirm that no single baseline dominates all cities and metrics. Crossformer, GWNet, MTGNN, and DLinear each obtain the best results on different cit--metric pairs, while performance varies more strongly on OSM+(UTD19) than on conventional saturated datasets. Extreme MAPE values are often associated with low-flow periods, heterogeneous city-level distributions, or sensor-to-road alignment noise. These phenomena are part of the intended stress-test value of OSM+: they reveal scalability and robustness bottlenecks that are less visible on small traffic benchmarks.

\subsection{Traffic Policy Control}

\begin{figure}[t]
    \centering
    \includegraphics[width=\linewidth]{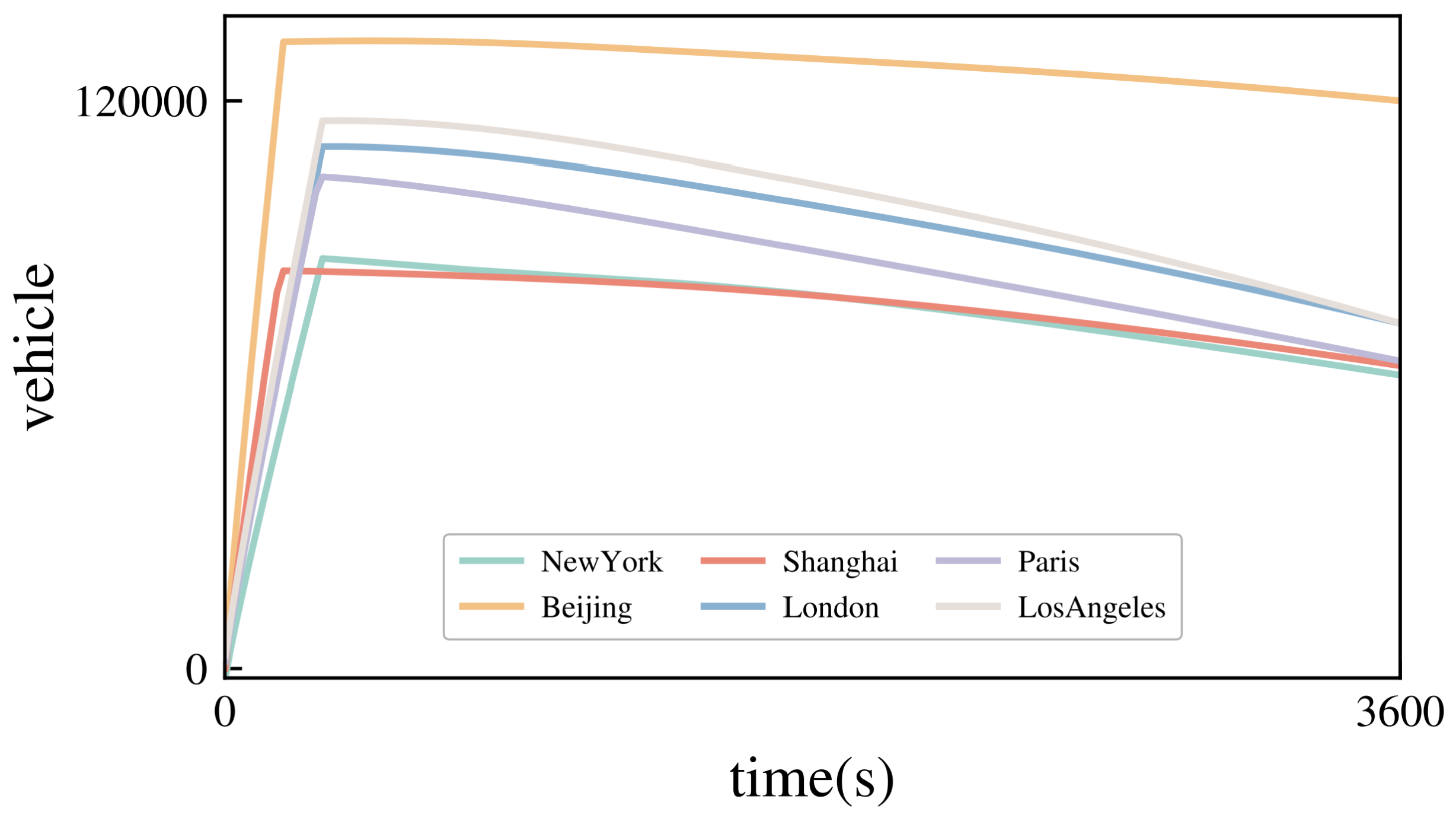}
    \caption{Active-vehicle curves of the six released traffic-control simulation scenarios under the generated demand profile.}
    \label{fig:tsc_curve}
\end{figure}

Starting from the cleaned OSM+ road network, we construct microscopic simulation environments for traffic policy control experiments, such as traffic signal control. As summarized in Figure~\ref{fig:framework}, the pipeline converts OSM+ road graphs into simulator-ready road networks, generates traffic demand, and configures the corresponding simulation files. This process enables OSM+ to serve not only as a static road-network database, but also as a source for building executable traffic-control benchmarks.

Compared with previously used datasets for traffic signal control experiments~\cite{wei2019survey}, our pipeline supports scenarios with more extensive spatial coverage and more realistic city-scale network structures. In this work, we release six representative city scenarios that have been carefully cleaned and validated. Table~\ref{traffic signal control} reports their basic statistics, including the number of signalized intersections and simulated vehicles. Figure~\ref{fig:tsc_curve} further shows the active-vehicle curves of the six released scenarios under the generated demand profile. The curves describe how the number of vehicles simultaneously present in the simulator changes over time. All scenarios exhibit a rapid increase in active vehicles at the beginning of the simulation, followed by a gradual decrease as vehicles complete their trips. The OSM+ based scenarios involve substantially larger vehicle populations than the smaller previous benchmarks, demonstrating that the released environments can impose heavier traffic loads and better stress-test controller scalability. These results indicate that the generated demands are able to create non-trivial and diverse city-scale traffic-control settings.

\section{Related Works}

\textbf{Map databases and services.} Map databases and services are the base for various urban research. The major map databases include two types: commercial services and open-source maps. The commercial services are provided by major vendors like Google~\citep{Google}, Bing~\citep{Bing}, Apple~\citep{Apple}, Baidu~\citep{Baidu}, and Gaode~\citep{Gaode}. Though varying in serving regions, these map providers usually only provide high-level APIs (e.g., plotting, routing) for users to develop their own applications. These kind of APIs cannot support research need for querying external elements on the map, e.g., population count. To facilitate faster research progress and small business, some open-source map providers release their products, including OpenStreetMap~\citep{osm1,osm2,osm3}, Mapbox~\citep{Mapbox}, Leaflet~\citep{Leaflet} and GeoServer~\citep{GeoServer}. These products store the map in various formats, which are difficult to be processed into a uniform format and need to be further cleaned for research use. 

\textbf{Map computing engine.} To better utilize map data, many map computing engines have been proposed. These tools can be categorized into two types: commercial map tools and open-source map tools. For \textit{commercial map engines}, \citet{googleapi} developed an online map application based on the Google Maps API, using commercial databases to provide users with complex data manipulation functions. \citet{meisam} processed remote sensing map data of Canadian agriculture on the Google Earth Engine (GEE) and obtained the annual crop list of Canada. \citet{uyen} combined GEE with the automatic water extraction index (AWEI) to address the long processing times in monitoring water surface dynamics. For \textit{open-source map engines}, \citet{wiam} accomplished the process of converting large-scale databases collected by cars into road tracks. \citet{boeing2017osmnx} developed OSMnx, which simplifies data collection and road network analysis from the perspectives of graph theory, transportation, and urban design. Besides, existing OpenStreetMap data processing tools also include Osmosis~\citep{Osmosis}, Nominatim~\citep{Nominatim} and OSRM~\citep{OSRM}. 

Although these tools are able to process map data, they still suffer from two key limitations. First, the processed outputs are often not well-defined, uniform, or structured, and are typically tailored for one-time use. As a result, similar tasks require repetitive computation. Second, these tools usually run on local machines, limiting their capabilities to support large-scale analyses, such as global-level processing. The comparison of existing OpenStreetMap processing tools and OSM+ is shown in Table \ref{related_works_comparison}.

\begin{table}[htbp]
  \centering
  \caption{Comparison of OSM data processing tools.}
  \label{related_works_comparison}
  \resizebox{.96\linewidth}{!}{
  \begin{tabular}{c|c|c}
    \toprule
    \textbf{Method} & \textbf{Efficient Querying} & \textbf{Graph Structure Output} \\
    \midrule
    Overpass API & \ding{55} & \ding{55} \\
    Osmosis & \ding{55} & \ding{55} \\
    OSMnx        & \ding{55} & \ding{51} \\
    Nominatim & \ding{51} & \ding{55} \\
    OSRM & \ding{51} & \ding{55} \\
    \textbf{OSM+} & \ding{51} & \ding{51} \\
    \bottomrule
  \end{tabular}
  }
\end{table}

\section{Conclusion}

In this paper, we release OSM+ as a scalable data infrastructure for city-wide graph learning and urban computing. By transforming raw OpenStreetMap records into a structured, worldwide road-network graph with cloud-supported query and conversion tools, OSM+ lowers the barrier to using large-scale geospatial data in reproducible research. Beyond releasing billion-level graph tables, we show that OSM+ can serve as a common spatial backbone for diverse downstream studies, including functional city-boundary analysis, cross-city traffic forecasting, and large-scale traffic-control simulation. The new 31-city traffic prediction benchmark and six city-scale control scenarios further reveal that existing models still face substantial challenges in robustness, memory efficiency, and scalability when moving from small curated datasets to realistic urban networks. We hope that OSM+ will support future research on graph learning, transportation intelligence, and multimodal geospatial foundation models, while encouraging the community to build more generalizable and scalable methods for understanding and optimizing cities.

\newpage

\section*{Acknowledgment}
This work was sponsored by 2025 Shanghai Key Technology Research and Development Program, Next-Generation Information Technology Project under Grant No. 25511103800.

\section*{Impact Statement}
\textbf{Positive impact.} OSM+ provides an open, graph-structured, billion-scale road-network resource and accompanying query/conversion utilities built from OpenStreetMap. It lowers the barrier for reproducible geospatial and transportation research: instead of repeatedly cleaning raw OSM exports and re-implementing fragile preprocessing pipelines, practitioners can directly access a unified road-graph representation, efficient spatial queries, and converters for tasks such as urban analytics, traffic forecasting, simulation-based policy evaluation, and multimodal geospatial model training.

\textbf{Potential negative impact.} Because OpenStreetMap is crowdsourced, OSM+ inherits regional variation in coverage and annotation quality. Models or benchmarks built on top of OSM+ may perform better in well-mapped regions and generalize poorly to sparsely mapped regions if this variation is ignored. In traffic forecasting, the same issue can also increase sensor-to-road matching noise when a sensor located on a missing minor road is projected to a more distant major road. Users should therefore report coverage, stratify evaluation by data quality when possible, preserve uncertainty in downstream analyses, and combine OSM+ with complementary local data sources for operational deployments. The UTD19-based traffic benchmark should be interpreted as a reproducible offline research benchmark rather than an online traffic-management ground truth.

\bibliography{references}
\bibliographystyle{icml2026}

\newpage
\appendix

\onecolumn

\section*{Appendix}

\section{Data fields} \label{data field}
In total, we release six core tables: \texttt{osm\_node}, \texttt{osm\_split\_edge}, \texttt{osm\_fulltag\_edge}, \texttt{osm\_node\_roadnet}, \texttt{osm\_split\_edge\_roadnet}, and \texttt{osm\_fulltag\_edge\_roadnet}. The first three are full OSM+ tables, which retain both road and non-road OpenStreetMap records, including POI nodes, building-related objects, administrative objects, and their associated tags. The latter three are roadnet-only tables, which are filtered from the full tables and are mainly used for road-network analysis and the experiments in this paper. This separation also clarifies the scope of our release: OSM+ provides the processed OpenStreetMap tables constructed by our pipeline, while external datasets used in downstream benchmarks remain subject to their original licenses and access policies.

\textbf{Full tables.} These three tables \texttt{osm\_node}, \texttt{osm\_split\_edge}, and \texttt{osm\_fulltag\_edge} contain the processed full OpenStreetMap snapshot information, including road and non-road records. Nodes and edges contain tag information; we parse frequently occurring node and edge attributes into structured fields and preserve the original OSM identifiers to support downstream joins.

\textbf{Roadnet-only tables.} \texttt{osm\_node\_roadnet}, \texttt{osm\_split\_edge\_roadnet} and \texttt{osm\_fulltag\_edge\_roadnet} contain roadnet-only information. These three roadnet-only tables contain only the nodes and edges on the extracted road network, rather than all OpenStreetMap objects. They describe the road-network graph used in our experiments. 

The field descriptions for the six OSM+ tables are as follows:

\renewcommand{\arraystretch}{1.2}
\begin{longtable}{p{0.2\linewidth} p{0.75\linewidth}}
\caption{The field descriptions of \textbf{\textsc{table}} \texttt{osm\_node}.} \\
\hline
\textbf{Field} & \textbf{Description} \\
\hline
\endfirsthead

\hline
\textbf{Field} & \textbf{Description} \\
\hline
\endhead

\hline
\endfoot

\texttt{osmid} & The unique OpenStreetMap identifier of the node. \\
\hline
\texttt{x} & Indicates the longitude of the node. \\
\hline
\texttt{y} & Indicates the latitude of the node. \\
\hline
\texttt{continent} & Indicates the continent, country, or broad geographic region associated with the node in this dataset. \\
\hline
\texttt{name} & Indicates the name assigned to the OSM feature. \\
\hline
\texttt{aerialway} & The type of aerial transport feature, such as a cable car, chair lift, gondola, or drag lift. \\
\hline
\texttt{aeroway} & The type of aviation-related feature, such as an airport, runway, taxiway, helipad, or terminal. \\
\hline
\texttt{amenity} & The type of useful facility or service for visitors or residents, such as a school, hospital, restaurant, parking area, or place of worship. \\
\hline
\texttt{barrier} & The type of physical barrier or obstacle, such as a fence, wall, gate, bollard, hedge, or block. \\
\hline
\texttt{boundary} & The type of boundary feature, such as an administrative boundary, protected area boundary, or political boundary. \\
\hline
\texttt{admin\_level} & Indicates the administrative hierarchy level of an administrative boundary. \\
\hline
\texttt{building} & The type of building or built structure represented by the node. \\
\hline
\texttt{entrance} & The type or function of an entrance to a building, site, or enclosed area. \\
\hline
\texttt{height} & Indicates the physical height of the feature, usually expressed in metres unless another unit is specified. \\
\hline
\texttt{craft} & The type of craft, trade, or specialized manual-service workplace. \\
\hline
\texttt{emergency} & The type of emergency-related facility, access point, or equipment. \\
\hline
\texttt{geological} & The type of geological feature or formation, such as an outcrop, moraine, or fault. \\
\hline
\texttt{healthcare} & The type of healthcare facility or healthcare-related service. \\
\hline
\texttt{highway} & The type of road, street, path, track, pedestrian way, or other transport route. \\
\hline
\texttt{ford} & Indicates whether the feature is a ford, where a road or path crosses a waterway at water level. \\
\hline
\texttt{lit} & Indicates whether the feature is lit by artificial lighting. \\
\hline
\texttt{historic} & The type of historic feature, such as a monument, archaeological site, memorial, castle, ruins, or historic building. \\
\hline
\texttt{landuse} & The type of human land use, such as residential, commercial, industrial, farmland, forest, or recreation ground. \\
\hline
\texttt{leisure} & The type of leisure, recreation, or sports-related facility or area. \\
\hline
\texttt{man\_made} & The type of human-made structure or object, such as a tower, pier, mast, pipeline, silo, or water tower. \\
\hline
\texttt{military} & The type of military facility, area, installation, or object. \\
\hline
\texttt{natural} & The type of natural or physical geographic feature, such as water, wood, peak, beach, cliff, wetland, or grassland. \\
\hline
\texttt{office} & The type of office or administrative workplace. \\
\hline
\texttt{place} & The type of named place or settlement, such as a country, state, city, town, village, hamlet, suburb, locality, or island. \\
\hline
\texttt{population} & Indicates the population associated with a place or administrative area. \\
\hline
\texttt{is\_in} & Indicates the larger geographic or administrative area in which the feature is located. \\
\hline
\texttt{power} & The type of power-generation, transmission, or distribution feature. \\
\hline
\texttt{public\_transport} & The type or role of public-transport feature, such as a stop position, platform, station, or stop area. \\
\hline
\texttt{railway} & The type of railway-related feature, such as a rail line, station, halt, tram, subway, or level crossing. \\
\hline
\texttt{shop} & The type of retail shop or commercial outlet. \\
\hline
\texttt{sport} & The type of sport or physical activity associated with the feature. \\
\hline
\texttt{telecom} & The type of telecommunications-related feature, such as a communication tower, exchange, antenna, or data centre. \\
\hline
\texttt{tourism} & The type of tourism-related feature, such as an attraction, hotel, museum, viewpoint, campsite, or information point. \\
\hline
\texttt{water} & The type of water body or water-related area. \\
\hline
\texttt{waterway} & The type of waterway feature, such as a river, stream, canal, drain, ditch, dam, weir, or lock. \\
\hline
\texttt{crossing} & The type or characteristics of a road, railway, or pedestrian crossing. \\
\hline

\end{longtable}

\renewcommand{\arraystretch}{1.2}
\begin{longtable}{p{0.2\linewidth} p{0.75\linewidth}}
\caption{The field descriptions of \textbf{\textsc{table}} \texttt{osm\_fulltag\_edge}.} \\
\hline
\textbf{Field} & \textbf{Description} \\
\hline
\endfirsthead

\hline
\textbf{Field} & \textbf{Description} \\
\hline
\endhead

\hline
\endfoot

\texttt{osmid} & The unique OpenStreetMap identifier of the edge or way. \\
\hline
\texttt{continent} & Indicates the continent, country, or broad geographic region associated with the edge in this dataset. \\
\hline
\texttt{name} & Indicates the name assigned to the OSM feature, road, way, or place. \\
\hline
\texttt{aerialway} & The type of aerial transport feature, such as a cable car, chair lift, gondola, or drag lift. \\
\hline
\texttt{aeroway} & The type of aviation-related feature, such as a runway, taxiway, airport, helipad, apron, or terminal. \\
\hline
\texttt{amenity} & The type of useful facility or service for visitors or residents, such as a school, hospital, restaurant, parking area, or place of worship. \\
\hline
\texttt{barrier} & The type of physical barrier or obstacle, such as a fence, wall, gate, bollard, hedge, or block. \\
\hline
\texttt{boundary} & The type of boundary feature, such as an administrative boundary, protected area boundary, or political boundary. \\
\hline
\texttt{admin\_level} & Indicates the administrative hierarchy level of an administrative boundary. \\
\hline
\texttt{building} & The type of building or built structure represented by the feature. \\
\hline
\texttt{entrance} & The type or function of an entrance to a building, site, or enclosed area. \\
\hline
\texttt{craft} & The type of craft, trade, or specialized manual-service workplace. \\
\hline
\texttt{emergency} & The type of emergency-related facility, access point, or equipment. \\
\hline
\texttt{geological} & The type of geological feature or formation, such as an outcrop, moraine, or fault. \\
\hline
\texttt{healthcare} & The type of healthcare facility or healthcare-related service. \\
\hline
\texttt{highway} & The functional type of road, street, path, track, pedestrian way, or other road-transport route. \\
\hline
\texttt{footway} & The type or role of pedestrian way, such as sidewalk, crossing, access aisle, or traffic island. \\
\hline
\texttt{sidewalk} & Indicates the presence, position, or type of sidewalk associated with the road. \\
\hline
\texttt{cycleway} & The type of cycling infrastructure associated with the way, such as a cycle lane, track, shared lane, or opposite-direction cycleway. \\
\hline
\texttt{abutters} & Indicates the land use or character of properties bordering the road or way. \\
\hline
\texttt{bus\_bay} & Indicates the presence or position of a bus bay beside the road. \\
\hline
\texttt{junction} & The type of road junction, such as a roundabout, circular junction, or motorway junction. \\
\hline
\texttt{lanes} & Indicates the number of traffic lanes available on the road or way. \\
\hline
\texttt{lit} & Indicates whether the road or feature is lit by artificial lighting. \\
\hline
\texttt{motorroad} & Indicates whether the road has motor-road status or motorway-like legal restrictions. \\
\hline
\texttt{oneway} & Indicates whether traffic is legally permitted in only one direction along the way. \\
\hline
\texttt{overtaking} & Indicates whether overtaking is legally allowed, restricted, or prohibited on the road. \\
\hline
\texttt{priority\_road} & Indicates whether the road has priority-road status. \\
\hline
\texttt{service} & The type of service road, service track, or service function, such as driveway, alley, parking aisle, siding, or yard track. \\
\hline
\texttt{smoothness} & Indicates the physical usability or surface quality of the way for wheeled vehicles. \\
\hline
\texttt{surface} & The type of surface material or finish of the way, such as asphalt, concrete, gravel, dirt, paving stones, or grass. \\
\hline
\texttt{turn} & Indicates turning direction or turn-lane information associated with the way. \\
\hline
\texttt{historic} & The type of historic feature, such as a monument, archaeological site, memorial, castle, ruins, or historic building. \\
\hline
\texttt{landuse} & The type of human land use, such as residential, commercial, industrial, farmland, forest, or recreation ground. \\
\hline
\texttt{leisure} & The type of leisure, recreation, or sports-related facility or area. \\
\hline
\texttt{man\_made} & The type of human-made structure or object, such as a tower, pier, mast, pipeline, silo, or water tower. \\
\hline
\texttt{military} & The type of military facility, area, installation, or object. \\
\hline
\texttt{natural} & The type of natural or physical geographic feature, such as water, wood, peak, beach, cliff, wetland, or grassland. \\
\hline
\texttt{office} & The type of office or administrative workplace. \\
\hline
\texttt{place} & The type of named place or settlement, such as a country, state, city, town, village, hamlet, suburb, locality, or island. \\
\hline
\texttt{population} & Indicates the population associated with a place or administrative area. \\
\hline
\texttt{is\_in} & Indicates the larger geographic or administrative area in which the feature is located. \\
\hline
\texttt{power} & The type of power-generation, transmission, or distribution feature. \\
\hline
\texttt{public\_transport} & The type or role of public-transport feature, such as a stop position, platform, station, or stop area. \\
\hline
\texttt{railway} & The type of railway-related feature, such as a rail line, station, halt, tram, subway, or level crossing. \\
\hline
\texttt{usage} & Indicates the usage or importance class of railway or transport infrastructure, such as main, branch, industrial, military, or tourism use. \\
\hline
\texttt{route} & The type of route represented by or associated with the feature, such as road, bus, bicycle, hiking, railway, or ferry route. \\
\hline
\texttt{shop} & The type of retail shop or commercial outlet. \\
\hline
\texttt{sport} & The type of sport or physical activity associated with the feature. \\
\hline
\texttt{telecom} & The type of telecommunications-related feature, such as a communication tower, exchange, antenna, or data centre. \\
\hline
\texttt{tourism} & The type of tourism-related feature, such as an attraction, hotel, museum, viewpoint, campsite, or information point. \\
\hline
\texttt{water} & The type of water body or water-related area. \\
\hline
\texttt{waterway} & The type of waterway feature, such as a river, stream, canal, drain, ditch, dam, weir, or lock. \\
\hline
\texttt{area} & Indicates whether a closed way represents an area rather than only a linear feature. \\
\hline
\texttt{bridge} & Indicates whether the way is on a bridge, and may specify the bridge type. \\
\hline
\texttt{access} & Indicates the general legal access permission for the feature, such as yes, no, private, permissive, or destination. \\
\hline
\texttt{est\_width} & Indicates the estimated width of the way when an exact measured width is not available. \\
\hline
\texttt{maxwidth} & Indicates the maximum legal or physical width permitted for vehicles using the way. \\
\hline
\texttt{maxaxleload} & Indicates the maximum permitted axle load for vehicles using the way. \\
\hline
\texttt{maxheight} & Indicates the maximum legal or physical vehicle height permitted on the way. \\
\hline
\texttt{maxlength} & Indicates the maximum legal or physical vehicle length permitted on the way. \\
\hline
\texttt{maxstay} & Indicates the maximum allowed duration of stay, commonly used for parking or stopping restrictions. \\
\hline
\texttt{maxweight} & Indicates the maximum permitted total vehicle weight on the way. \\
\hline
\texttt{maxspeed} & Indicates the maximum legal speed limit on the way. \\
\hline
\texttt{minspeed} & Indicates the minimum legal speed required on the way. \\
\hline

\end{longtable}

\renewcommand{\arraystretch}{1.2}
\begin{longtable}{p{0.2\linewidth} p{0.75\linewidth}}
\caption{The field descriptions of \textbf{\textsc{table}} \texttt{osm\_split\_edge}.} \\
\hline
\textbf{Field} & \textbf{Description} \\
\hline
\endfirsthead

\hline
\textbf{Field} & \textbf{Description} \\
\hline
\endhead

\hline
\endfoot

\texttt{lsid} & The unique identifier of a split edge segment. \\
\hline
\texttt{osmid\_start} & The OpenStreetMap identifier of the starting node of a split edge segment. \\
\hline
\texttt{osmid\_start\_x} & The longitude of the starting node of a split edge segment. \\
\hline
\texttt{osmid\_start\_y} & The latitude of the starting node of a split edge segment. \\
\hline
\texttt{osmid\_end} & The OpenStreetMap identifier of the ending node of a split edge segment. \\
\hline
\texttt{osmid\_end\_x} & The longitude of the ending node of a split edge segment. \\
\hline
\texttt{osmid\_end\_y} & The latitude of the ending node of a split edge segment. \\
\hline

\end{longtable}

\renewcommand{\arraystretch}{1.2}
\begin{longtable}{p{0.2\linewidth} p{0.75\linewidth}}
\caption{The field descriptions of \textbf{\textsc{table}} \texttt{osm\_node\_roadnet}.} \\
\hline
\textbf{Field} & \textbf{Description} \\
\hline
\endfirsthead

\hline
\textbf{Field} & \textbf{Description} \\
\hline
\endhead

\hline
\endfoot

\texttt{osmid} & A unique identifier for each roadnet node. \\
\hline
\texttt{x} & The longitude of a roadnet node. \\
\hline
\texttt{y} & The latitude of a roadnet node. \\
\hline
\end{longtable}

\renewcommand{\arraystretch}{1.2}
\begin{longtable}{p{0.2\linewidth} p{0.75\linewidth}}
\caption{The field descriptions of \textbf{\textsc{table}} \texttt{osm\_fulltag\_edge\_roadnet}.} \\
\hline
\textbf{Field} & \textbf{Description} \\
\hline
\endfirsthead

\hline
\textbf{Field} & \textbf{Description} \\
\hline
\endhead

\hline
\endfoot

\texttt{osmid} & The unique OpenStreetMap identifier of one road. \\
\hline
\texttt{highway} & The road hierarchy for different roads, starting from the highest level: motorway, trunk, primary, secondary, etc. \\
\hline

\end{longtable}

\renewcommand{\arraystretch}{1.2}
\begin{longtable}{p{0.2\linewidth} p{0.75\linewidth}}
\caption{The field descriptions of \textbf{\textsc{table}} \texttt{osm\_split\_edge}.} \\
\hline
\textbf{Field} & \textbf{Description} \\
\hline
\endfirsthead

\hline
\textbf{Field} & \textbf{Description} \\
\hline
\endhead

\hline
\endfoot

\texttt{lsid} & The unique identifier of a split road segment. \\
\hline
\texttt{osmid\_start} & The OpenStreetMap identifier of the starting node of a split road segment. \\
\hline
\texttt{osmid\_start\_x} & The longitude of the starting node of a split road segment. \\
\hline
\texttt{osmid\_start\_y} & The latitude of the starting node of a split road segment. \\
\hline
\texttt{osmid\_end} & The OpenStreetMap identifier of the ending node of a split road segment. \\
\hline
\texttt{osmid\_end\_x} & The longitude of the ending node of a split road segment. \\
\hline
\texttt{osmid\_end\_y} & The latitude of the ending node of a split road segment. \\
\hline

\end{longtable}
\section{Data and Code Access} \label{data_access}


Our OSM+ dataset is made available under the \textbf{Open Database License (ODbL 1.0)}: \url{http://opendatacommons.org/licenses/odbl/1.0/}. Any rights in individual contents of the database are licensed under the Database Contents License: \url{http://opendatacommons.org/licenses/dbcl/1.0/}. These tables are released as standard downloadable files through \url{https://huggingface.co/datasets/suziyang/OSM_plus}, allowing users to download complete tables or region-level shards for local processing.

\noindent\textbf{Preprocessing pipeline.} The complete data preprocessing pipeline is released at \url{https://github.com/SJTU-CILAB/OSM-dataset}. External users can follow the step-by-step tutorial to reproduce the data construction process from raw Geofabrik mirrors, including converting raw OSM files into global csv files and generating the corresponding OSM+ tables. All codes in this repository are released under \textbf{MIT License}.

\textbf{Efficient query tools.} We provide query tools based on the \textbf{City Brain Platform }(\url{https://workspace.citybrain.org/#/home}) to support efficient querying and management of OSM+ data. The operational workflows and scripts for using this platform with OSM+ are available at \url{https://github.com/SJTU-CILAB/OSM-dataset}. The query interface is useful for extracting city-scale subgraphs, issuing SQL-style spatial queries, and joining road segments with auxiliary tags without downloading all global-scale tables.

\textbf{Benchmark code and data.} All experimental configurations, source code, processed benchmark graphs, and converters for the downstream tasks in this paper are released at \url{https://github.com/SJTU-CILAB/OSM-dataset}. These resources support reproduction of the city-boundary detection, traffic-prediction, and traffic-policy-control experiments.

\textbf{Third-party data boundary.} The downstream benchmarks released in this paper integrate OSM+ with third-party upstream data sources such as UTD19~\citep{loder2020utd19} , GADM data~\citep{GADM}, and nightlight data. These third-party sources remain subject to their original licenses, access policies, and redistribution terms. We release the alignment code and processed benchmark structures to support reproducibility, but we do not claim ownership of third-party upstream data.

\section{Dataset Maintenance} \label{maintenance}

The OSM+ release evaluated in this paper is a static snapshot processed from Geofabrik/OpenStreetMap data downloaded in \textbf{March 2026}, and this fixed snapshot is used for all reported experiments. Since OpenStreetMap is continuously edited by volunteers, a single static copy can become stale over time; therefore, we maintain OSM+ using a time-based versioning strategy and plan to update the dataset every three months. Each release will be named by its release month, e.g., \texttt{OSM+ vYYYY.MM}, and archived independently so that researchers can reproduce a specific benchmark snapshot while also accessing newer road-network states for current applications. In addition, the released preprocessing pipeline allows users to process a newer Geofabrik\texttt{.osm.pbf} file for a particular region of interest without waiting for the next official global release.

\section{Preprocessing Pipeline}

The overall procedure of the OpenStreetMap data processing can be described as the following steps:
\begin{enumerate}
    \item Download continental OSM files from \url{http://download.geofabrik.de/}. The snapshot used for the experiments in this paper is the March 2026 Geofabrik/OpenStreetMap snapshot.
    \item Decompose continental OSM files (larger than 100GB) into several smaller OSM files.
    \item Convert OSM files into CSV files, including \textbf{nodes.csv} and \textbf{edges.csv}. In node files, each record represents a node on OpenStreetMap with about 30 attributes. In edge files, each record represents an edge-like OSM object.
    \item Split linestrings on OpenStreetMap and transform \textbf{edges.csv} files into start--end pairs. Each edge file is transformed into a \textbf{fulltag.csv} file and a \textbf{split.csv} file. The schema of \textbf{fulltag.csv} is the same as \textbf{edges.csv} except for removing the ``nodes'' tag. Each record in \textbf{split.csv} represents a link on OpenStreetMap.
    \item Upload these files to ODPS/MaxCompute for distributed cleaning and joining.
    \item Select nodes and edges on the roadnet from tables on ODPS. Finally, three roadnet-only tables and three full tables are generated. Detailed schemas are introduced in Appendix~\ref{data field}.
\end{enumerate}

\section{Auxiliary Attribute Completeness} \label{aux-completeness}

OSM+ parses auxiliary OSM tags, including road hierarchy and POI-related attributes, into structured fields. These attributes are useful for downstream tasks, but their completeness naturally varies across regions because OSM is maintained by local contributors. Well-mapped regions often provide dense POIs and consistent road hierarchy tags, whereas sparsely mapped regions may contain many generic labels such as \texttt{unclassified} and fewer POI annotations. We preserve this original distribution rather than interpolating missing attributes. Users who rely on auxiliary attributes should report attribute coverage, such as the fraction of road edges with valid hierarchy tags and POI density per square kilometer, before drawing cross-region conclusions.
\section{Detailed Experiment Settings}
\subsection{Basic Query Operation}
This section describes the details of optimizing basic query operations. For point query, we divide the entire map into grids of 0.2° by 0.2° in latitude and longitude, and determine which grid the point to be queried is located in. Then, we only need to retrieve data points in adjacent grids during retrieval. For the nearest neighbor query, we also used the "window check" method. In addition, we utilized the k-d tree, a binary tree that represents a division of k-dimensional space to reduce the time complexity.

\subsection{City Boundary Detection}
This section mainly describes the process of using OSM+ data for city boundary detection. First, we select several representative areas for experiment, including the China Yangtze River Delta (115°E$\sim$122°E, 29°N$\sim$32°N), New York in the USA (84°W$\sim$70.5°W, 37.8°N$\sim$45.6°N), Central Europe (2°W$\sim$11.5°E, 45.5°N$\sim$53°N), and Nigeria in Africa (0.5°E$\sim$14°E, 4.2°N$\sim$11.7°N).  We integrate the roadnet data, raster image population data, and raster image nightlight data of these areas by converting them into point-wise data. Next, we perform clustering on the road net data, nightlight data, and population data respectively, and then weigh and sum the results of clustering to obtain a new clustering result. After that, we divide the entire map into a grid structure with a size of 0.05° in latitude and longitude. We then count the number of points of each type within each grid and assign the grid to the category with the most points, obtaining a rough boundary. Finally, we examine the empty grids by considering a grid nine times its size centered on it and reclassifying it to make the boundary smoother. This process allows us to discover a more reasonable city boundary.

\subsection{City-scale Traffic Prediction} \label{traffic prediction details}

We conduct all the experiments on machines with two NVIDIA RTX 3090 GPUs (24GB VRAM each) and 128 GB of CPU memory on Ubuntu 20.04. All models are implemented in Python 3. For the problem setting, both the input and output sequence lengths are set to 12. We use traffic-flow data from UTD19~\citep{loder2020utd19} to construct a reproducible offline benchmark, rather than an online deployment setting or an authoritative ground truth for operational traffic management. For all 31 city datasets, the training, validation, and test sets are split at a ratio of 6:2:2. We evaluate performance using mean absolute error (MAE), root mean squared error (RMSE), and mean absolute percentage error (MAPE).

\textbf{Sensor-to-road matching.} For each UTD19 city, we first extract the corresponding OSM+ road-network subgraph by spatial range. We then align UTD19 sensor or measurement coordinates to OSM+ through nearest-edge matching: each observation location is projected to the closest physical road segment in the OSM+ roadnet, and the associated road-network topology is used to construct the prediction graph. This matching procedure is deterministic and scalable, but it may introduce localization drift when OSM coverage is sparse or when the true minor road is missing from OSM. A global quantitative matching-error audit would require lane-level ground truth or high-frequency trajectories for all cities, which are not available across the full benchmark. We therefore preserve the matching noise and report it as part of the realistic cross-city benchmark challenge.

Here we list some details of our implemented baseline methods. For hyperparameters that are not mentioned, we adopt the default hyperparameters from BasicTS~\citep{shao2023exploring}: 

\textbf{AGCRN}~\citep{bai2020adaptive}. We use two layers of AGCRN to capture the node-specific spatial and temporal dynamics. For the hyperparameters, we set the hidden unit to 64 for all the AGCRN cells and the batch size to 64. We set the learning rate to 0.003 and the embedding dimension to 64 for all 31 city datasets extracted from the OSM+ dataset. Besides, we choose $L_1$ Loss as the loss function.

\textbf{Crossformer}~\citep{zhang2022crossformer}. When implementing Crossformer, we set segment length $L_{seg}$ to 24, as it is related to both the model performance and computation efficiency. Besides, we set the window size to 2. We use Adam optimizer with a 0.0002 learning rate and 0.0005 weight decay rate.

\textbf{DCRNN}~\citep{li2017diffusion}. Both encoder and decoder contain two recurrent layers. In each recurrent layer, there are 64 units and the initial learning rate is set to 0.003. Besides, the maximum step of random walks, i.e., $K$, is set to 2. For scheduled sampling, the thresholded inverse sigmoid function is used as the probability decay:
\begin{equation}
    \epsilon_i=\frac{\tau}{\tau+\exp (i / \tau)}
\end{equation}
where $i$ is the number of iterations while $\tau$ is the parameter to control the speed of convergence, which is set to 2,000 in the experiments.

\textbf{DLinear}~\citep{diagne2012lyapunov}. For implementation details about DLinear, we adopt the default hyperparameters from ~\citep{shao2023exploring} to train the models. The training epoch is set to 100.

\textbf{FEDformer}~\citep{zhou2022fedformer}. The FEDformer is trained using the Adam optimizer with a learning rate of 0.0005. The batch size is set to 64. An early stopping counter is employed to stop the training process after three epochs if no loss degradation on the validation set is observed.

\textbf{GWNet}~\citep{wu2019graph}. We use two layers of Graph WaveNet with a sequence of dilation factors $\{1, 2\}$. We randomly initialize node embeddings by a uniform distribution with a size of 10. We train our model using the Adam optimizer with an initial learning rate of 0.0005. Dropout with $p=0.3$ is applied to the outputs of the graph convolution layer. 

\subsection{Traffic Policy Control}
We use the script provided by CBData~\citep{liang2023cblab} to convert the OSM+ roadnet file to the corresponding format required by CBEngine. We then conduct traffic-flow simulation experiments in six cities (Beijing, London, Los Angeles, New York, Paris, and Shanghai). The simulation runs for 3600 steps, with traffic demand injected uniformly during the first 300 steps. The released scenario files include the converted road network, signal phase configuration, generated demand, and scripts for computing standard control metrics such as average travel time, queue length, and delay.

The released scenarios are designed to support comparable controller evaluation. Fixed-time, max-pressure, and learning-based controllers can be evaluated using the same roadnet, demand files, and metric scripts, ensuring that future methods can be compared using identical city-scale inputs.



\section{Extended Experiment Results} \label{extended results}

To further illustrate the supporting performance of the OSM+ dataset over traffic flow prediction task, we repeat each set of experiments five times and report their means and standard deviations on horizon 3, 6 and 12, following the common setting in this problem~\citep{tedjopurnomo2020survey}. Comprehensive results are shown in Table \ref{tab: horizon3}, Table \ref{tab: horizon6} and Table \ref{tab: horizon12}. The correspondence between city names and their abbreviations is shown as follows: AGB (Augsburg), BSL (Basel), BRN (Bern), BHX (Birmingham), BOL (Bolton), BOD (Bordeaux), BRE (Bremen), KN (Constance), DA (Darmstadt), ESS (Essen), FRA (Frankfurt), GRZ (Graz), GRQ (Groningen), HAM (Hamburg), INN (Innsbruck), KS (Kassel), MAN (Manchester), MEL (Melbourne), RTM (Rotterdam), SDR (Santander), SP (Speyer), SXB (Strasbourg), STR (Stuttgart), TPE (Taipeh/Taipei), TO (Torino), YTO (Toronto), TLS (Toulouse), UTC (Utrecht), VNO (Vilnius), WOB (Wolfsburg), and ZRH (Zurich).

\begin{small}
\setlength{\tabcolsep}{0.6mm}
\renewcommand{\arraystretch}{1.4}
\begin{longtable}{c|c|c|c|c|c|c|c|c}
\caption{The performance comparison for seven baseline methods over 31 real-world city datasets with \textbf{horizon=3}. The best results in each row are in bold. All experiments are repeated five times, and the mean and standard deviation are reported.}
\label{tab: horizon3}\\

\toprule
     City & Metric & AGCRN & Crossformer & DCRNN & DLinear & FEDformer & GWNet & MTGNN \\ \hline
     \multirow{3}{*}{AGB} & MAE & 42.53 $\pm$ 0.22 & \textbf{39.89 $\pm$ 0.11} & OOM & 40.56 $\pm$ 0.00 & 46.15 $\pm$ 0.03 & 40.65 $\pm$ 0.19 & 40.76 $\pm$ 0.38 \\ 
     & RMSE & 87.82 $\pm$ 0.22 & 85.21 $\pm$ 0.92 & OOM & 82.00 $\pm$ 0.00 & 99.68 $\pm$ 0.34 & \textbf{81.14 $\pm$ 1.17} & 83.04 $\pm$ 0.33 \\ 
     & MAPE(\%) & 38.97 $\pm$ 0.18 & \textbf{31.21 $\pm$ 1.72} & OOM & 32.43 $\pm$ 0.02 & 39.76 $\pm$ 0.10 & 32.09 $\pm$ 0.95 & 33.27 $\pm$ 2.39 \\ \hline
     \multirow{3}{*}{BSL} & MAE & 55.00 $\pm$ 0.15 & 49.76 $\pm$ 0.96 & 116.48 $\pm$ 1.50 & \textbf{49.12 $\pm$ 0.00} & 52.55 $\pm$ 0.18 & 68.47 $\pm$ 0.15 & 65.86 $\pm$ 0.60 \\ 
     & RMSE & 82.77 $\pm$ 0.08 & \textbf{74.24 $\pm$ 2.83} & 155.13 $\pm$ 0.43 & 79.39 $\pm$ 0.03 & 76.74 $\pm$ 0.37 & 93.83 $\pm$ 1.64 & 90.52 $\pm$ 1.20 \\ 
     & MAPE(\%) & 45.85 $\pm$ 0.35 & \textbf{43.38 $\pm$ 2.60} & 171.29 $\pm$ 7.33 & 42.93 $\pm$ 0.01 & 52.67 $\pm$ 1.15 & 91.08 $\pm$ 3.34 & 75.48 $\pm$ 2.24 \\ \hline
      \multirow{3}{*}{BRN} & MAE & 46.82 $\pm$ 1.30 & \textbf{45.65 $\pm$ 0.17} & OOM & 48.76 $\pm$ 0.02 & 51.16 $\pm$ 0.10 & 47.49 $\pm$ 0.56 & 59.36 $\pm$ 0.86 \\ 
     & RMSE & 468.78 $\pm$ 9.95 & \textbf{454.56 $\pm$ 1.64} & OOM & 454.98 $\pm$ 0.18 & 431.93 $\pm$ 0.45 & 440.49 $\pm$ 7.76 & 477.32 $\pm$ 13.68 \\ 
     & MAPE(\%) & \textbf{222.50 $\pm$ 13.23} & 233.32 $\pm$ 18.35 & OOM & 229.36 $\pm$ 2.05 & 232.49 $\pm$ 17.64 & 278.14 $\pm$ 19.92 & 347.74 $\pm$ 20.37 \\ \hline
     \multirow{3}{*}{BHX} & MAE & 88.84 $\pm$ 0.35 & 84.86 $\pm$ 7.57 & 289.70 $\pm$ 11.26 & 92.86 $\pm$ 0.80 & 104.66 $\pm$ 0.44 & 90.95 $\pm$ 0.18 & \textbf{84.13 $\pm$ 0.64} \\ 
     & RMSE & 137.05 $\pm$ 0.61 & 134.05 $\pm$ 8.50 & 425.16 $\pm$ 37.43 & 144.99 $\pm$ 0.88 & 158.84 $\pm$ 0.19 & 144.59 $\pm$ 1.12 & \textbf{130.56 $\pm$ 0.88} \\ 
     & MAPE(\%) & 43.38 $\pm$ 0.35 & \textbf{40.51 $\pm$ 2.58} & 191.60 $\pm$ 57.29 & 48.95 $\pm$ 0.11 & 52.14 $\pm$ 0.90 & 51.86 $\pm$ 1.83 & 45.67 $\pm$ 3.76 \\ \hline
     \multirow{3}{*}{BOL} & MAE & \textbf{29.63 $\pm$ 0.27} & 32.03 $\pm$ 1.73 & 32.74 $\pm$ 0.38 & 33.31 $\pm$ 0.02 & 34.19 $\pm$ 0.04 & 32.69 $\pm$ 1.92 & 29.68 $\pm$ 0.23 \\ 
     & RMSE & 84.46 $\pm$ 0.07 & 84.23 $\pm$ 0.25 & 87.89 $\pm$ 0.04 & 83.81 $\pm$ 0.06 & 87.51 $\pm$ 0.13 & 84.37 $\pm$ 0.56 & \textbf{81.28 $\pm$ 0.13} \\
     & MAPE(\%) & \textbf{18.44 $\pm$ 1.09} & 19.69 $\pm$ 0.44 & 22.48 $\pm$ 0.69 & 30.71 $\pm$ 0.10 & 26.86 $\pm$ 1.25 & 22.51 $\pm$ 0.53 & 18.63 $\pm$ 0.50 \\ \hline
     \multirow{3}{*}{BOD} & MAE & 62.34 $\pm$ 0.22 & 60.21 $\pm$ 0.60 & 244.58 $\pm$ 20.30 & \textbf{58.23 $\pm$ 0.00} & 58.51 $\pm$ 0.30 & 59.78 $\pm$ 0.05 & 71.17 $\pm$ 1.11 \\ 
     & RMSE & 100.52 $\pm$ 0.47 & 95.41 $\pm$ 0.16 & 324.45 $\pm$ 24.77 & 92.86 $\pm$ 0.01 & \textbf{92.09 $\pm$ 0.27} & 93.00 $\pm$ 0.09 & 109.95 $\pm$ 2.32 \\ 
     & MAPE(\%) & 37.45 $\pm$ 1.21 & \textbf{33.47 $\pm$ 0.43} & 280.19 $\pm$ 14.03 & 37.59 $\pm$ 0.05 & 38.28 $\pm$ 0.84 & 43.06 $\pm$ 0.33 & 45.75 $\pm$ 0.42 \\ \hline
     \multirow{3}{*}{BRE} & MAE & 55.40 $\pm$ 0.02 & 56.33 $\pm$ 1.40 & OOM & 58.99 $\pm$ 0.03 & 58.80 $\pm$ 0.04 & 55.45 $\pm$ 0.09 & \textbf{55.18 $\pm$ 0.17} \\     
     & RMSE & 92.18 $\pm$ 0.08 & 93.06 $\pm$ 1.74 & OOM & 97.06 $\pm$ 0.11 & 95.33 $\pm$ 0.02 & \textbf{91.31 $\pm$ 0.11} & 91.72 $\pm$ 0.48 \\ 
     & MAPE(\%) & 36.08 $\pm$ 0.26 & \textbf{33.69 $\pm$ 1.66} & OOM & 40.12 $\pm$ 0.12 & 40.41 $\pm$ 0.49 & 36.10 $\pm$ 0.27 & 34.93 $\pm$ 0.22 \\ \hline
     \multirow{3}{*}{KN} & MAE & OOM & 39.17 $\pm$ 1.36 & 119.00 $\pm$ 0.09 & \textbf{35.47 $\pm$ 0.00} & 38.35 $\pm$ 0.37 & 37.55 $\pm$ 0.55 & 40.06 $\pm$ 0.38 \\ 
     & RMSE & OOM & 62.75 $\pm$ 2.15 & 152.84 $\pm$ 0.67 & \textbf{54.99 $\pm$ 0.00} & 58.78 $\pm$ 0.32 & 58.21 $\pm$ 0.67 & 61.15 $\pm$ 0.86 \\ 
     & MAPE(\%) & OOM & 47.64 $\pm$ 0.56 & 293.93 $\pm$ 0.88 & 54.02 $\pm$ 0.17 & 61.77 $\pm$ 1.36 & \textbf{47.54 $\pm$ 0.59} & 58.71 $\pm$ 1.08 \\ \hline
     \multirow{3}{*}{DA} & MAE & 56.32 $\pm$ 0.04 & 51.12 $\pm$ 0.79 & OOM & \textbf{50.77 $\pm$ 0.00} & 54.26 $\pm$ 0.08 & 51.80 $\pm$ 0.11 & 54.58 $\pm$ 0.23 \\ 
     & RMSE & 88.65 $\pm$ 0.20 & 77.01 $\pm$ 1.12 & OOM & \textbf{75.41 $\pm$ 0.00} & 79.78 $\pm$ 0.25 & 78.39 $\pm$ 0.16 & 86.54 $\pm$ 1.08 \\ 
     & MAPE(\%) & 50.79 $\pm$ 0.00 & \textbf{45.51 $\pm$ 0.49} & OOM & 51.38 $\pm$ 0.02 & 58.16 $\pm$ 0.89 & 48.27 $\pm$ 0.64 & 48.10 $\pm$ 0.11 \\ \hline
     \multirow{3}{*}{ESS} & MAE & 40.66 $\pm$ 0.28 & 36.95 $\pm$ 0.10 & 172.66 $\pm$ 0.19 & 41.06 $\pm$ 0.00 & 40.51 $\pm$ 0.07 & 36.86 $\pm$ 0.08 & \textbf{35.94 $\pm$ 0.07} \\ 
     & RMSE & 59.07 $\pm$ 0.76 & 53.29 $\pm$ 0.27 & 224.47 $\pm$ 3.72 & 59.39 $\pm$ 0.01 & 58.41 $\pm$ 0.03 & 53.72 $\pm$ 0.22 & \textbf{52.57 $\pm$ 0.03} \\   
     & MAPE(\%) & 35.54 $\pm$ 1.10 & 34.79 $\pm$ 1.28 & 302.81 $\pm$ 15.18 & 34.62 $\pm$ 0.40 & 36.01 $\pm$ 0.71 & 32.85 $\pm$ 0.28 & \textbf{31.51 $\pm$ 2.29} \\ \hline      
     \multirow{3}{*}{FRA} & MAE & 124.69 $\pm$ 0.05 & 140.18 $\pm$ 14.57 & 185.87 $\pm$ 2.29 & \textbf{71.58 $\pm$ 0.30} & 84.83 $\pm$ 1.13 & 139.06 $\pm$ 1.84 & 170.44 $\pm$ 12.29 \\ 
     & RMSE & 157.09 $\pm$ 0.09 & 166.53 $\pm$ 16.07 & 237.20 $\pm$ 2.82 & \textbf{90.90 $\pm$ 1.09} & 108.23 $\pm$ 2.00 & 163.25 $\pm$ 2.09 & 190.32 $\pm$ 12.38 \\ 
     & MAPE(\%) & 39.40 $\pm$ 0.02 & 41.18 $\pm$ 3.19 & 47.65 $\pm$ 0.59 & \textbf{20.21 $\pm$ 0.07} & 23.41 $\pm$ 0.15 & 42.48 $\pm$ 0.68 & 51.70 $\pm$ 3.62 \\ \hline
     \multirow{3}{*}{GRZ} & MAE & 59.06 $\pm$ 0.02 & \textbf{50.55 $\pm$ 1.02} & 174.96 $\pm$ 0.00 & 52.27 $\pm$ 0.00 & 50.95 $\pm$ 0.24 & 55.07 $\pm$ 0.74 & 53.48 $\pm$ 0.62 \\ 
     & RMSE & 89.07 $\pm$ 0.02 & 74.28 $\pm$ 0.84 & 220.49 $\pm$ 0.00 & 75.95 $\pm$ 0.01 & \textbf{74.23 $\pm$ 0.17} & 81.10 $\pm$ 1.00 & 82.35 $\pm$ 1.49 \\ 
     & MAPE(\%) & 116.57 $\pm$ 0.80 & 68.30 $\pm$ 4.29 & 467.17 $\pm$ 0.00 & \textbf{63.68 $\pm$ 0.15} & 66.96 $\pm$ 2.12 & 68.11 $\pm$ 1.24 & 71.41 $\pm$ 0.40 \\ \hline
     \multirow{3}{*}{GRQ} & MAE & 63.24 $\pm$ 0.01 & 62.25 $\pm$ 1.53 & 162.87 $\pm$ 0.00 & \textbf{60.08 $\pm$ 0.41} & 72.11 $\pm$ 1.47 & 64.33 $\pm$ 0.16 & 63.21 $\pm$ 1.50 \\ 
     & RMSE & 87.02 $\pm$ 0.08 & 86.03 $\pm$ 0.07 & 218.69 $\pm$ 0.00 & \textbf{82.72 $\pm$ 0.56} & 99.96 $\pm$ 1.73 & 88.41 $\pm$ 0.03 & 86.27 $\pm$ 1.51 \\ 
     & MAPE(\%) & 30.46 $\pm$ 0.19 & \textbf{28.82 $\pm$ 0.24} & 112.24 $\pm$ 0.00 & 31.41 $\pm$ 0.29 & 36.83 $\pm$ 0.57 & 31.79 $\pm$ 0.19 & 30.87 $\pm$ 1.08 \\ \hline
     \multirow{3}{*}{HAM} & MAE & 45.68 $\pm$ 0.09 & 43.50 $\pm$ 0.10 & 97.51 $\pm$ 0.56 & 44.82 $\pm$ 0.00 & 46.34 $\pm$ 0.05 & \textbf{43.28 $\pm$ 0.01} & 44.05 $\pm$ 0.09 \\ 
     & RMSE & 75.36 $\pm$ 0.60 & 70.76 $\pm$ 0.19 & 150.67 $\pm$ 2.84 & 73.13 $\pm$ 0.01 & 74.96 $\pm$ 0.18 & \textbf{70.69 $\pm$ 0.07} & 73.02 $\pm$ 0.19 \\ 
     & MAPE(\%) & 44.67 $\pm$ 0.36 & \textbf{42.79 $\pm$ 0.51} & 111.42 $\pm$ 5.09 & 47.93 $\pm$ 0.10 & 48.93 $\pm$ 0.17 & 43.48 $\pm$ 0.16 & 43.30 $\pm$ 0.13 \\ \hline
     \multirow{3}{*}{INN} & MAE & 70.05 $\pm$ 0.15 & 67.34 $\pm$ 1.24 & 333.14 $\pm$ 2.31 & 76.43 $\pm$ 0.01 & 70.66 $\pm$ 0.65 & \textbf{65.65 $\pm$ 0.12} & OOM \\ 
     & RMSE & 101.84 $\pm$ 0.05 & 97.03 $\pm$ 0.57 & 443.91 $\pm$ 2.79 & 113.93 $\pm$ 0.09 & 102.62 $\pm$ 0.80 & \textbf{95.28 $\pm$ 0.17} & OOM \\ 
     & MAPE(\%) & 31.38 $\pm$ 0.83 & 35.24 $\pm$ 5.44 & 304.36 $\pm$ 0.76 & 33.84 $\pm$ 0.45 & 30.70 $\pm$ 0.10 & \textbf{28.00 $\pm$ 0.42} & OOM \\ \hline
     \multirow{3}{*}{KS} & MAE & 69.88 $\pm$ 0.37 & 73.22 $\pm$ 1.27 & 244.28 $\pm$ 0.00 & \textbf{63.22 $\pm$ 2.52} & 80.77 $\pm$ 0.65 & 63.34 $\pm$ 1.06 & 154.46 $\pm$ 1.90 \\ 
     & RMSE & 203.93 $\pm$ 0.33 & 204.17 $\pm$ 2.19 & 342.80 $\pm$ 0.00 & 158.72 $\pm$ 4.21 & 177.02 $\pm$ 0.31 & \textbf{155.08 $\pm$ 0.53} & 233.89 $\pm$ 3.73 \\ 
     & MAPE(\%) & \textbf{80.45 $\pm$ 1.04} & 87.42 $\pm$ 3.53 & 440.12 $\pm$ 0.00 & 81.40 $\pm$ 4.85 & 105.62 $\pm$ 1.65 & 81.47 $\pm$ 2.36 & 235.91 $\pm$ 0.42 \\ \hline
     \multirow{3}{*}{MAN} & MAE & 97.96 $\pm$ 0.71 & 87.80 $\pm$ 0.65 & 336.72 $\pm$ 0.00 & 92.36 $\pm$ 5.65 & 99.05 $\pm$ 1.29 & \textbf{84.26 $\pm$ 0.20} & 85.24 $\pm$ 0.93 \\ 
     & RMSE & 169.67 $\pm$ 0.73 & 160.48 $\pm$ 1.62 & 448.89 $\pm$ 0.00 & 156.98 $\pm$ 7.85 & 167.30 $\pm$ 2.46 & \textbf{151.48 $\pm$ 0.49} & 154.15 $\pm$ 2.57 \\ 
     & MAPE(\%) & 39.66 $\pm$ 1.35 & 37.73 $\pm$ 0.54 & 283.06 $\pm$ 0.00 & 42.24 $\pm$ 2.20 & 46.21 $\pm$ 1.60 & \textbf{32.83 $\pm$ 0.26} & 36.21 $\pm$ 1.39 \\ \hline
     \multirow{3}{*}{MEL} & MAE & 37.39 $\pm$ 0.00 & 36.21 $\pm$ 0.95 & OOM & 42.57 $\pm$ 0.85 & 40.80 $\pm$ 0.58 & 36.26 $\pm$ 0.01 & \textbf{35.86 $\pm$ 0.95} \\      
     & RMSE & 56.63 $\pm$ 0.01 & \textbf{54.30 $\pm$ 2.21} & OOM & 64.49 $\pm$ 1.03 & 60.80 $\pm$ 1.36 & 54.35 $\pm$ 0.06 & 54.78 $\pm$ 1.68 \\ 
     & MAPE(\%) & 37.60 $\pm$ 0.03 & 35.63 $\pm$ 0.72 & OOM & 37.82 $\pm$ 0.99 & 43.56 $\pm$ 0.04 & \textbf{27.22 $\pm$ 0.07} & 31.77 $\pm$ 0.80 \\ \hline
     \multirow{3}{*}{RTM} & MAE & \textbf{49.08 $\pm$ 0.16} & 50.46 $\pm$ 0.06 & 170.03 $\pm$ 0.00 & 55.93 $\pm$ 0.01 & 58.45 $\pm$ 0.73 & 54.45 $\pm$ 0.28 & 51.45 $\pm$ 0.08 \\ 
     & RMSE & \textbf{87.20 $\pm$ 0.12} & 87.80 $\pm$ 0.13 & 232.29 $\pm$ 0.00 & 95.05 $\pm$ 0.02 & 96.26 $\pm$ 0.90 & 92.68 $\pm$ 0.52 & 88.92 $\pm$ 0.36 \\ 
     & MAPE(\%) & \textbf{36.76 $\pm$ 0.05} & 41.26 $\pm$ 0.01 & 320.02 $\pm$ 0.00 & 44.61 $\pm$ 0.19 & 57.44 $\pm$ 1.81 & 43.72 $\pm$ 1.04 & 38.46 $\pm$ 1.26 \\ \hline
     \multirow{3}{*}{SDR} & MAE & 88.07 $\pm$ 0.25 & 86.61 $\pm$ 1.76 & 259.05 $\pm$ 0.00 & 80.68 $\pm$ 0.02 & 107.62 $\pm$ 1.33 & \textbf{77.84 $\pm$ 0.41} & 78.50 $\pm$ 2.26 \\ 
     & RMSE & 230.51 $\pm$ 0.94 & 231.19 $\pm$ 0.49 & 434.97 $\pm$ 0.00 & \textbf{187.89 $\pm$ 0.01} & 216.83 $\pm$ 2.05 & 202.47 $\pm$ 1.86 & 211.89 $\pm$ 0.85 \\ 
     & MAPE(\%) & 54.00 $\pm$ 1.15 & 53.10 $\pm$ 6.85 & 257.82 $\pm$ 0.00 & 43.89 $\pm$ 0.16 & 81.61 $\pm$ 0.53 & 44.22 $\pm$ 3.89 & \textbf{36.34 $\pm$ 0.17} \\ \hline
     \multirow{3}{*}{SP} & MAE & 48.92 $\pm$ 0.05 & \textbf{47.51 $\pm$ 0.03} & 121.12 $\pm$ 2.04 & 50.74 $\pm$ 0.01 & 52.37 $\pm$ 0.11 & 47.84 $\pm$ 0.05 & 47.75 $\pm$ 0.11 \\ 
     & RMSE & 70.45 $\pm$ 0.06 & \textbf{68.34 $\pm$ 0.07} & 172.55 $\pm$ 9.56 & 72.60 $\pm$ 0.06 & 74.68 $\pm$ 0.06 & 68.91 $\pm$ 0.17 & 68.90 $\pm$ 0.02 \\ 
     & MAPE(\%) & 39.84 $\pm$ 0.53 & \textbf{36.77 $\pm$ 0.11} & 102.07 $\pm$ 25.06 & 43.49 $\pm$ 0.00 & 43.79 $\pm$ 0.59 & 37.92 $\pm$ 0.33 & 37.04 $\pm$ 0.13 \\ \hline
     \multirow{3}{*}{SXB} & MAE & 76.69 $\pm$ 0.07 & 73.93 $\pm$ 0.14 & 259.67 $\pm$ 0.00 & 78.84 $\pm$ 0.05 & 80.54 $\pm$ 0.24 & 74.23 $\pm$ 0.03 & \textbf{73.87 $\pm$ 0.41} \\      
     & RMSE & 134.85 $\pm$ 0.14 & \textbf{130.71 $\pm$ 0.32} & 360.44 $\pm$ 0.00 & 138.48 $\pm$ 0.03 & 141.88 $\pm$ 0.15 & 131.69 $\pm$ 0.05 & 131.14 $\pm$ 0.23 \\ 
     & MAPE(\%) & 40.04 $\pm$ 1.09 & 40.73 $\pm$ 2.09 & 223.03 $\pm$ 0.00 & 43.50 $\pm$ 0.18 & 43.51 $\pm$ 0.74 & 37.64 $\pm$ 0.26 & \textbf{36.53 $\pm$ 0.63} \\ \hline
     \multirow{3}{*}{STR} & MAE & 57.40 $\pm$ 0.43 & 57.84 $\pm$ 3.60 & 59.86 $\pm$ 1.67 & 60.07 $\pm$ 0.08 & 68.33 $\pm$ 1.18 & \textbf{55.70 $\pm$ 0.02} & OOM \\ 
     & RMSE & 74.79 $\pm$ 0.28 & 75.55 $\pm$ 4.24 & 78.25 $\pm$ 3.02 & 78.53 $\pm$ 0.16 & 89.06 $\pm$ 1.27 & \textbf{72.13 $\pm$ 0.07} & OOM \\ 
     & MAPE(\%) & 18.44 $\pm$ 0.08 & 19.11 $\pm$ 2.18 & 19.07 $\pm$ 0.70 & 20.01 $\pm$ 0.01 & 22.32 $\pm$ 0.45 & \textbf{18.10 $\pm$ 0.06} & OOM \\ \hline
     \multirow{3}{*}{TPE} & MAE & 126.81 $\pm$ 0.07 & 126.02 $\pm$ 3.81 & 490.25 $\pm$ 8.98 & 125.41 $\pm$ 0.01 & 134.61 $\pm$ 0.09 & \textbf{117.30 $\pm$ 0.40} & 121.00 $\pm$ 0.90 \\ 
     & RMSE & 555.59 $\pm$ 1.23 & 561.67 $\pm$ 10.18 & 988.24 $\pm$ 11.03 & \textbf{482.59 $\pm$ 0.38} & 540.48 $\pm$ 5.15 & 493.02 $\pm$ 0.48 & 512.90 $\pm$ 1.36 \\ 
     & MAPE(\%) & 42.20 $\pm$ 0.35 & 45.74 $\pm$ 5.09 & 266.56 $\pm$ 4.39 & 41.80 $\pm$ 0.12 & 47.72 $\pm$ 0.55 & \textbf{38.30 $\pm$ 1.75} & 40.01 $\pm$ 1.54 \\ \hline
     \multirow{3}{*}{TO} & MAE & 77.75 $\pm$ 0.14 & 71.06 $\pm$ 0.96 & 313.88 $\pm$ 0.21 & \textbf{68.52 $\pm$ 0.01} & 74.48 $\pm$ 0.07 & 74.91 $\pm$ 1.05 & 80.07 $\pm$ 1.05 \\ 
     & RMSE & 131.86 $\pm$ 0.13 & 114.97 $\pm$ 2.10 & 415.51 $\pm$ 1.03 & \textbf{111.00 $\pm$ 0.07} & 118.37 $\pm$ 0.35 & 116.63 $\pm$ 0.82 & 124.72 $\pm$ 1.31 \\ 
     & MAPE(\%) & 50.86 $\pm$ 0.32 & \textbf{39.42 $\pm$ 0.29} & 400.19 $\pm$ 7.19 & 40.27 $\pm$ 0.47 & 48.37 $\pm$ 0.98 & 46.33 $\pm$ 2.97 & 52.21 $\pm$ 2.04 \\ \hline
     \multirow{3}{*}{YTO} & MAE & 45.81 $\pm$ 0.12 & 45.25 $\pm$ 0.40 & 129.55 $\pm$ 18.89 & 59.49 $\pm$ 0.01 & 52.16 $\pm$ 0.04 & 46.68 $\pm$ 0.25 & \textbf{44.65 $\pm$ 0.07} \\
     & RMSE & 77.29 $\pm$ 0.30 & \textbf{74.48 $\pm$ 0.06} & 192.41 $\pm$ 21.36 & 95.84 $\pm$ 0.09 & 81.85 $\pm$ 0.04 & 77.47 $\pm$ 0.47 & 75.03 $\pm$ 0.00 \\ 
     & MAPE(\%) & 33.49 $\pm$ 0.02 & \textbf{32.72 $\pm$ 1.55} & 114.20 $\pm$ 27.77 & 42.70 $\pm$ 0.66 & 46.45 $\pm$ 1.69 & 35.18 $\pm$ 0.38 & 33.36 $\pm$ 2.07 \\ \hline
     \multirow{3}{*}{TLS} & MAE & 257.32 $\pm$ 0.36 & \textbf{255.21 $\pm$ 0.28} & 264.24 $\pm$ 6.15 & 263.84 $\pm$ 0.00 & 296.07 $\pm$ 0.02 & 255.32 $\pm$ 0.00 & 259.00 $\pm$ 0.03 \\ 
     & RMSE & 349.78 $\pm$ 0.86 & 342.51 $\pm$ 0.43 & 352.70 $\pm$ 9.62 & 348.15 $\pm$ 0.03 & 410.70 $\pm$ 0.70 & \textbf{341.22 $\pm$ 0.34} & 349.09 $\pm$ 0.86 \\ 
     & MAPE(\%) & 754.88 $\pm$ 3.75 & 746.38 $\pm$ 4.66 & 791.79 $\pm$ 77.77 & 872.83 $\pm$ 0.91 & 836.22 $\pm$ 1.06 & 747.55 $\pm$ 5.53 & \textbf{726.09 $\pm$ 9.80} \\ \hline
     \multirow{3}{*}{UTC} & MAE & OOM & 51.67 $\pm$ 0.80 & OOM & 44.40 $\pm$ 1.21 & 61.74 $\pm$ 1.37 & 75.85 $\pm$ 0.25 & \textbf{39.61 $\pm$ 0.20} \\      
     & RMSE & OOM & 83.26 $\pm$ 3.39 & OOM & 75.24 $\pm$ 1.49 & 91.06 $\pm$ 0.40 & 120.97 $\pm$ 0.66 & \textbf{68.41 $\pm$ 0.37} \\      
     & MAPE(\%) & OOM & 57.21 $\pm$ 15.30 & OOM & 45.37 $\pm$ 1.44 & 77.81 $\pm$ 4.77 & 91.81 $\pm$ 5.44 & \textbf{37.55 $\pm$ 2.65} \\ \hline      
     \multirow{3}{*}{VNO} & MAE & 82.75 $\pm$ 0.08 & 77.98 $\pm$ 0.27 & OOM & \textbf{69.08 $\pm$ 1.53} & 81.57 $\pm$ 0.28 & 69.25 $\pm$ 0.23 & 89.80 $\pm$ 1.00 \\ 
     & RMSE & 112.24 $\pm$ 0.29 & 106.39 $\pm$ 0.56 & OOM & \textbf{94.18 $\pm$ 2.13} & 110.65 $\pm$ 0.82 & 94.42 $\pm$ 0.06 & 118.34 $\pm$ 1.00 \\ 
     & MAPE(\%) & 47.89 $\pm$ 0.26 & 42.38 $\pm$ 0.14 & OOM & 35.59 $\pm$ 0.75 & 42.38 $\pm$ 0.60 & \textbf{35.32 $\pm$ 0.81} & 57.14 $\pm$ 1.75 \\ \hline
     \multirow{3}{*}{WOB} & MAE & 52.66 $\pm$ 0.05 & 51.73 $\pm$ 1.69 & 54.51 $\pm$ 0.05 & 56.05 $\pm$ 0.02 & 54.89 $\pm$ 0.01 & 51.27 $\pm$ 0.18 & \textbf{51.12 $\pm$ 0.01} \\
     & RMSE & 81.81 $\pm$ 0.31 & 79.60 $\pm$ 3.58 & 84.03 $\pm$ 0.30 & 85.65 $\pm$ 0.02 & 82.31 $\pm$ 0.02 & \textbf{78.15 $\pm$ 0.24} & 78.27 $\pm$ 0.04 \\ 
     & MAPE(\%) & 40.76 $\pm$ 0.38 & \textbf{37.38 $\pm$ 0.17} & 42.23 $\pm$ 0.24 & 46.71 $\pm$ 0.06 & 46.64 $\pm$ 0.10 & 39.56 $\pm$ 0.58 & 39.41 $\pm$ 0.80 \\ \hline
     \multirow{3}{*}{ZRH} & MAE & OOM & 53.23 $\pm$ 0.05 & OOM & 56.33 $\pm$ 0.00 & 56.72 $\pm$ 0.08 & 59.24 $\pm$ 7.75 & \textbf{52.34 $\pm$ 0.18} \\
     & RMSE & OOM & 75.13 $\pm$ 0.05 & OOM & 79.32 $\pm$ 0.01 & 79.36 $\pm$ 0.01 & 83.96 $\pm$ 11.26 & \textbf{74.05 $\pm$ 0.34} \\      
     & MAPE(\%) & OOM & 35.77 $\pm$ 1.18 & OOM & 40.87 $\pm$ 0.04 & 41.31 $\pm$ 0.53 & 44.88 $\pm$ 11.14 & \textbf{34.95 $\pm$ 0.46} \\ \hline 
\bottomrule
\end{longtable}
\end{small}

\begin{small}

\setlength{\tabcolsep}{0.6mm}
\renewcommand{\arraystretch}{1.4}
\begin{longtable}{c|c|c|c|c|c|c|c|c}
\caption{The performance comparison for seven baseline methods over 31 real-world city datasets with \textbf{horizon=6}. The best results in each row are in bold. All experiments are repeated five times, and the mean and standard deviation are reported.}
\label{tab: horizon6}\\
\toprule
     City & Metric & AGCRN & Crossformer & DCRNN & DLinear & FEDformer & GWNet & MTGNN \\ \hline
         \multirow{3}{*}{AGB} & MAE & 47.27 $\pm$ 0.41 & \textbf{43.29 $\pm$ 0.11} & OOM & 46.25 $\pm$ 0.00 & 50.21 $\pm$ 0.02 & 46.22 $\pm$ 0.59 & 45.61 $\pm$ 0.55 \\ 
        & RMSE & 100.54 $\pm$ 0.22 & \textbf{95.50 $\pm$ 1.21} & OOM & 97.90 $\pm$ 0.03 & 112.57 $\pm$ 0.37 & 97.17 $\pm$ 2.27 & 97.30 $\pm$ 0.43 \\ 
        & MAPE(\%) & 43.35 $\pm$ 0.73 & \textbf{33.68 $\pm$ 0.45} & OOM & 36.33 $\pm$ 0.15 & 42.69 $\pm$ 0.09 & 36.93 $\pm$ 1.56 & 37.46 $\pm$ 3.09 \\ \hline
        \multirow{3}{*}{BSL} & MAE & 63.95 $\pm$ 0.02 & 61.81 $\pm$ 3.57 & 118.10 $\pm$ 0.84 & \textbf{59.01 $\pm$ 0.01} & 59.94 $\pm$ 1.42 & 81.55 $\pm$ 1.76 & 78.88 $\pm$ 1.13 \\ 
         & RMSE & 94.86 $\pm$ 0.09 & 91.78 $\pm$ 4.55 & 157.37 $\pm$ 1.12 & 99.52 $\pm$ 0.04 & \textbf{87.06 $\pm$ 2.86} & 114.99 $\pm$ 4.41 & 110.81 $\pm$ 2.49 \\ 
         & MAPE(\%) & 54.81 $\pm$ 0.20 & 68.11 $\pm$ 2.43 & 175.12 $\pm$ 9.70 & \textbf{49.76 $\pm$ 0.16} & 59.65 $\pm$ 0.08 & 109.78 $\pm$ 3.98 & 91.64 $\pm$ 2.08 \\ \hline
        \multirow{3}{*}{BRN} & MAE & 49.20 $\pm$ 1.52 & \textbf{48.47 $\pm$ 0.32} & OOM & 51.65 $\pm$ 0.01 & 54.01 $\pm$ 0.18 & 50.27 $\pm$ 0.10 & 69.27 $\pm$ 0.82 \\ 
         & RMSE & 481.50 $\pm$ 16.19 & 465.00 $\pm$ 1.10 & OOM & 464.66 $\pm$ 0.33 & \textbf{450.70 $\pm$ 0.37} & 468.88 $\pm$ 0.39 & 508.09 $\pm$ 12.22 \\ 
         & MAPE(\%) & 216.94 $\pm$ 6.34 & \textbf{186.02 $\pm$ 20.85} & OOM & 244.39 $\pm$ 5.93 & 242.43 $\pm$ 19.54 & 293.55 $\pm$ 20.22 & 411.51 $\pm$ 30.03 \\ \hline
        \multirow{3}{*}{BHX} & MAE & 101.87 $\pm$ 0.00 & \textbf{86.82 $\pm$ 7.11} & 299.18 $\pm$ 30.69 & 103.14 $\pm$ 2.28 & 113.03 $\pm$ 0.16 & 99.17 $\pm$ 0.39 & 89.68 $\pm$ 0.34 \\ 
         & RMSE & 148.33 $\pm$ 0.23 & \textbf{133.22 $\pm$ 9.67} & 426.46 $\pm$ 61.57 & 156.29 $\pm$ 2.41 & 175.26 $\pm$ 0.51 & 150.67 $\pm$ 0.76 & 137.11 $\pm$ 1.22 \\ 
         & MAPE(\%) & 59.01 $\pm$ 0.56 & \textbf{45.71 $\pm$ 1.30} & 229.60 $\pm$ 44.64 & 57.28 $\pm$ 1.81 & 57.12 $\pm$ 0.60 & 56.87 $\pm$ 5.04 & 49.85 $\pm$ 1.96 \\ \hline
        \multirow{3}{*}{BOL} & MAE & \textbf{30.99 $\pm$ 0.09} & 32.53 $\pm$ 1.31 & 37.09 $\pm$ 0.61 & 35.99 $\pm$ 0.07 & 35.71 $\pm$ 0.38 & 33.47 $\pm$ 0.60 & 31.57 $\pm$ 0.32 \\ 
         & RMSE & 87.74 $\pm$ 0.14 & 87.00 $\pm$ 0.68 & 93.43 $\pm$ 0.01 & 88.23 $\pm$ 0.03 & 87.93 $\pm$ 0.81 & 82.40 $\pm$ 0.12 & \textbf{82.26 $\pm$ 0.01} \\         
         & MAPE(\%) & \textbf{19.58 $\pm$ 1.26} & 20.30 $\pm$ 0.26 & 26.38 $\pm$ 0.99 & 33.56 $\pm$ 0.24 & 29.66 $\pm$ 1.49 & 28.79 $\pm$ 0.72 & 21.63 $\pm$ 0.37 \\ \hline
        \multirow{3}{*}{BOD} & MAE & 69.25 $\pm$ 0.17 & 65.32 $\pm$ 0.92 & 242.05 $\pm$ 12.05 & \textbf{64.43 $\pm$ 0.01} & 64.68 $\pm$ 0.20 & 70.55 $\pm$ 0.22 & 86.45 $\pm$ 0.91 \\ 
         & RMSE & 112.27 $\pm$ 0.50 & 103.95 $\pm$ 0.50 & 321.50 $\pm$ 16.12 & 102.17 $\pm$ 0.01 & \textbf{101.21 $\pm$ 0.16} & 108.43 $\pm$ 0.09 & 135.49 $\pm$ 2.53 \\ 
         & MAPE(\%) & 39.52 $\pm$ 1.05 & \textbf{35.61 $\pm$ 0.49} & 280.80 $\pm$ 1.26 & 42.14 $\pm$ 0.23 & 42.19 $\pm$ 0.72 & 52.86 $\pm$ 0.88 & 55.08 $\pm$ 1.37 \\ \hline
        \multirow{3}{*}{BRE} & MAE & 56.30 $\pm$ 0.01 & 56.94 $\pm$ 0.75 & OOM & 62.32 $\pm$ 0.01 & 60.38 $\pm$ 0.08 & 56.86 $\pm$ 0.01 & \textbf{56.26 $\pm$ 0.27} \\          
        & RMSE & 93.86 $\pm$ 0.02 & 94.36 $\pm$ 0.99 & OOM & 101.83 $\pm$ 0.07 & 97.83 $\pm$ 0.11 & 93.62 $\pm$ 0.02 & \textbf{93.58 $\pm$ 0.67} \\          
        & MAPE(\%) & 36.30 $\pm$ 0.14 & \textbf{34.78 $\pm$ 0.07} & OOM & 41.95 $\pm$ 0.28 & 41.36 $\pm$ 0.48 & 37.02 $\pm$ 0.02 & 35.45 $\pm$ 0.29 \\ \hline
        \multirow{3}{*}{KN} & MAE & OOM & 43.36 $\pm$ 1.67 & 118.57 $\pm$ 2.65 & \textbf{37.85 $\pm$ 0.00} & 40.09 $\pm$ 0.24 & 43.01 $\pm$ 1.18 & 45.06 $\pm$ 0.30 \\ 
         & RMSE & OOM & 69.19 $\pm$ 2.16 & 151.98 $\pm$ 2.99 & 61.54 $\pm$ 0.01 & \textbf{61.47 $\pm$ 0.14} & 68.27 $\pm$ 1.68 & 68.58 $\pm$ 0.14 \\ 
         & MAPE(\%) & OOM & \textbf{47.64 $\pm$ 0.45} & 294.00 $\pm$ 8.92 & 59.41 $\pm$ 0.17 & 64.05 $\pm$ 1.65 & 53.57 $\pm$ 1.58 & 68.56 $\pm$ 3.07 \\ \hline
        \multirow{3}{*}{DA} & MAE & 58.34 $\pm$ 0.14 & \textbf{53.03 $\pm$ 0.05} & OOM & 53.93 $\pm$ 0.01 & 56.35 $\pm$ 0.07 & 55.08 $\pm$ 0.19 & 59.16 $\pm$ 0.15 \\ 
         & RMSE & 92.05 $\pm$ 0.44 & \textbf{80.86 $\pm$ 0.21} & OOM & 81.10 $\pm$ 0.07 & 83.79 $\pm$ 0.23 & 84.75 $\pm$ 0.59 & 96.79 $\pm$ 0.57 \\ 
         & MAPE(\%) & 51.28 $\pm$ 0.03 & \textbf{47.58 $\pm$ 0.26} & OOM & 53.46 $\pm$ 0.08 & 59.83 $\pm$ 0.81 & 50.41 $\pm$ 1.14 & 50.52 $\pm$ 0.23 \\ \hline
        \multirow{3}{*}{ESS} & MAE & 41.58 $\pm$ 0.11 & 39.69 $\pm$ 0.44 & 173.69 $\pm$ 0.75 & 47.27 $\pm$ 0.00 & 43.48 $\pm$ 0.01 & 38.23 $\pm$ 0.14 & \textbf{37.59 $\pm$ 0.06} \\
        & RMSE & 61.02 $\pm$ 0.33 & 57.73 $\pm$ 0.93 & 226.42 $\pm$ 2.28 & 68.86 $\pm$ 0.02 & 63.05 $\pm$ 0.26 & 56.05 $\pm$ 0.23 & \textbf{55.76 $\pm$ 0.09} \\          
        & MAPE(\%) & 34.29 $\pm$ 0.34 & 40.50 $\pm$ 4.51 & 299.82 $\pm$ 15.01 & 40.99 $\pm$ 0.15 & 39.62 $\pm$ 0.40 & 34.16 $\pm$ 0.17 & \textbf{34.01 $\pm$ 2.44} \\ \hline
        \multirow{3}{*}{FRA} & MAE & 158.46 $\pm$ 0.09 & 141.89 $\pm$ 33.64 & 187.73 $\pm$ 7.77 & \textbf{93.49 $\pm$ 1.42} & 107.62 $\pm$ 2.29 & 173.62 $\pm$ 3.08 & 258.38 $\pm$ 13.05 \\ 
         & RMSE & 192.87 $\pm$ 0.02 & 172.02 $\pm$ 35.30 & 236.87 $\pm$ 10.11 & \textbf{115.65 $\pm$ 1.17} & 136.75 $\pm$ 3.69 & 202.44 $\pm$ 2.63 & 279.20 $\pm$ 13.65 \\ 
         & MAPE(\%) & 52.05 $\pm$ 0.01 & 43.28 $\pm$ 8.91 & 52.30 $\pm$ 1.53 & \textbf{27.45 $\pm$ 0.40} & 30.92 $\pm$ 0.64 & 54.62 $\pm$ 1.16 & 81.61 $\pm$ 3.79 \\ \hline
        \multirow{3}{*}{GRZ} & MAE & 60.79 $\pm$ 0.01 & \textbf{52.70 $\pm$ 0.96} & 185.12 $\pm$ 0.00 & 58.29 $\pm$ 0.01 & 54.03 $\pm$ 0.24 & 58.32 $\pm$ 0.87 & 56.03 $\pm$ 0.20 \\ 
         & RMSE & 91.77 $\pm$ 0.08 & \textbf{77.53 $\pm$ 0.71} & 233.72 $\pm$ 0.00 & 84.30 $\pm$ 0.04 & 78.69 $\pm$ 0.12 & 86.47 $\pm$ 1.17 & 87.69 $\pm$ 0.61 \\ 
         & MAPE(\%) & 110.69 $\pm$ 1.77 & \textbf{65.44 $\pm$ 6.48} & 465.95 $\pm$ 0.00 & 72.50 $\pm$ 1.27 & 69.79 $\pm$ 3.62 & 71.75 $\pm$ 1.36 & 72.35 $\pm$ 0.12 \\ \hline
        \multirow{3}{*}{GRQ} & MAE & 68.43 $\pm$ 0.01 & 65.61 $\pm$ 1.26 & 161.01 $\pm$ 0.00 & \textbf{64.78 $\pm$ 0.91} & 79.46 $\pm$ 0.32 & 66.49 $\pm$ 0.22 & 71.04 $\pm$ 3.46 \\ 
         & RMSE & 93.25 $\pm$ 0.24 & 90.55 $\pm$ 0.25 & 217.62 $\pm$ 0.00 & \textbf{89.18 $\pm$ 1.54} & 110.72 $\pm$ 0.49 & 91.41 $\pm$ 0.15 & 95.64 $\pm$ 4.02 \\ 
         & MAPE(\%) & 33.78 $\pm$ 0.31 & \textbf{30.45 $\pm$ 1.73} & 114.12 $\pm$ 0.00 & 35.59 $\pm$ 0.64 & 41.81 $\pm$ 0.46 & 33.33 $\pm$ 0.42 & 36.21 $\pm$ 1.79 \\ \hline
        \multirow{3}{*}{HAM} & MAE & 46.44 $\pm$ 0.12 & 44.26 $\pm$ 0.07 & 97.38 $\pm$ 0.48 & 46.26 $\pm$ 0.01 & 47.54 $\pm$ 0.10 & \textbf{44.16 $\pm$ 0.01} & 45.04 $\pm$ 0.01 \\ 
         & RMSE & 77.82 $\pm$ 0.73 & 74.06 $\pm$ 0.10 & 150.66 $\pm$ 2.47 & 77.55 $\pm$ 0.01 & 79.25 $\pm$ 0.22 & \textbf{74.06 $\pm$ 0.10} & 79.08 $\pm$ 1.23 \\ 
         & MAPE(\%) & 45.57 $\pm$ 0.70 & 44.25 $\pm$ 2.04 & 111.18 $\pm$ 3.96 & 49.44 $\pm$ 0.06 & 50.07 $\pm$ 0.06 & \textbf{43.68 $\pm$ 0.20} & 44.06 $\pm$ 0.07 \\ \hline
        \multirow{3}{*}{INN} & MAE & 71.83 $\pm$ 0.48 & 67.78 $\pm$ 0.30 & 337.43 $\pm$ 6.70 & 86.47 $\pm$ 0.02 & 73.73 $\pm$ 0.55 & \textbf{66.95 $\pm$ 0.28} & OOM \\ 
         & RMSE & 104.72 $\pm$ 0.59 & 98.77 $\pm$ 1.04 & 452.25 $\pm$ 0.27 & 133.76 $\pm$ 0.11 & 107.19 $\pm$ 0.71 & \textbf{97.41 $\pm$ 0.45} & OOM \\ 
         & MAPE(\%) & 29.79 $\pm$ 1.13 & 34.04 $\pm$ 3.08 & 294.18 $\pm$ 24.63 & 37.85 $\pm$ 0.91 & 32.08 $\pm$ 0.02 & \textbf{28.82 $\pm$ 1.09} & OOM \\ \hline
        \multirow{3}{*}{KS} & MAE & 77.06 $\pm$ 0.53 & 83.40 $\pm$ 5.34 & 229.47 $\pm$ 0.00 & 72.68 $\pm$ 1.41 & 85.40 $\pm$ 0.05 & \textbf{70.79 $\pm$ 0.35} & 190.66 $\pm$ 2.64 \\ 
         & RMSE & 216.34 $\pm$ 0.41 & 218.00 $\pm$ 4.66 & 332.48 $\pm$ 0.00 & 177.06 $\pm$ 2.66 & 187.31 $\pm$ 0.67 & \textbf{174.04 $\pm$ 0.28} & 277.37 $\pm$ 5.51 \\ 
         & MAPE(\%) & 97.13 $\pm$ 1.41 & 111.00 $\pm$ 17.20 & 410.19 $\pm$ 0.00 & 99.24 $\pm$ 1.87 & 116.48 $\pm$ 0.61 & \textbf{93.42 $\pm$ 0.13} & 308.03 $\pm$ 0.63 \\ \hline
        \multirow{3}{*}{MAN} & MAE & 106.44 $\pm$ 1.13 & 94.31 $\pm$ 0.83 & 335.71 $\pm$ 0.00 & 103.64 $\pm$ 10.23 & 110.65 $\pm$ 0.92 & \textbf{93.95 $\pm$ 1.00} & 94.68 $\pm$ 2.13 \\ 
         & RMSE & 180.81 $\pm$ 1.54 & 169.06 $\pm$ 0.26 & 448.79 $\pm$ 0.00 & 172.31 $\pm$ 13.89 & 184.62 $\pm$ 2.27 & \textbf{163.00 $\pm$ 3.21} & 168.90 $\pm$ 4.47 \\ 
         & MAPE(\%) & 43.13 $\pm$ 1.40 & 43.34 $\pm$ 2.45 & 279.02 $\pm$ 0.00 & 47.57 $\pm$ 4.43 & 51.47 $\pm$ 0.57 & \textbf{37.68 $\pm$ 0.64} & 42.02 $\pm$ 1.16 \\ \hline
        \multirow{3}{*}{MEL} & MAE & 48.56 $\pm$ 0.00 & \textbf{43.98 $\pm$ 1.73} & OOM & 58.89 $\pm$ 0.02 & 50.66 $\pm$ 0.49 & 49.02 $\pm$ 0.20 & 46.12 $\pm$ 1.29 \\ 
         & RMSE & 75.34 $\pm$ 0.01 & \textbf{66.14 $\pm$ 3.55} & OOM & 89.81 $\pm$ 0.46 & 76.23 $\pm$ 0.70 & 74.99 $\pm$ 0.31 & 72.62 $\pm$ 2.27 \\ 
         & MAPE(\%) & 44.66 $\pm$ 0.05 & 40.38 $\pm$ 0.06 & OOM & 54.22 $\pm$ 0.50 & 53.21 $\pm$ 0.56 & \textbf{33.97 $\pm$ 0.06} & 39.97 $\pm$ 1.00 \\ \hline
        \multirow{3}{*}{RTM} & MAE & \textbf{51.36 $\pm$ 0.14} & 52.84 $\pm$ 1.05 & 179.34 $\pm$ 0.00 & 64.94 $\pm$ 0.05 & 65.13 $\pm$ 0.38 & 63.44 $\pm$ 0.77 & 55.79 $\pm$ 0.08 \\ 
         & RMSE & 92.19 $\pm$ 0.07 & \textbf{91.97 $\pm$ 0.65} & 240.67 $\pm$ 0.00 & 110.95 $\pm$ 0.04 & 106.66 $\pm$ 0.68 & 106.73 $\pm$ 1.16 & 97.29 $\pm$ 0.25 \\ 
         & MAPE(\%) & \textbf{39.55 $\pm$ 0.86} & 48.88 $\pm$ 3.29 & 349.98 $\pm$ 0.00 & 50.56 $\pm$ 0.21 & 63.69 $\pm$ 0.70 & 49.48 $\pm$ 0.75 & 40.87 $\pm$ 1.70 \\ \hline
        \multirow{3}{*}{SDR} & MAE & 101.26 $\pm$ 0.42 & 98.18 $\pm$ 2.76 & 256.61 $\pm$ 0.00 & 93.27 $\pm$ 0.06 & 119.78 $\pm$ 2.36 & \textbf{86.47 $\pm$ 0.70} & 91.34 $\pm$ 2.73 \\ 
         & RMSE & 252.60 $\pm$ 0.01 & 249.70 $\pm$ 1.91 & 433.47 $\pm$ 0.00 & \textbf{214.37 $\pm$ 0.03} & 239.59 $\pm$ 2.95 & 220.15 $\pm$ 1.88 & 235.42 $\pm$ 0.34 \\ 
         & MAPE(\%) & 58.36 $\pm$ 1.09 & 52.17 $\pm$ 5.47 & 255.29 $\pm$ 0.00 & 51.78 $\pm$ 0.42 & 89.07 $\pm$ 2.29 & 49.86 $\pm$ 5.91 & \textbf{41.60 $\pm$ 0.36} \\ \hline
         \multirow{3}{*}{SP} & MAE & 49.19 $\pm$ 0.05 & \textbf{47.78 $\pm$ 0.02} & 122.39 $\pm$ 4.19 & 52.40 $\pm$ 0.01 & 53.30 $\pm$ 0.09 & 48.28 $\pm$ 0.09 & 48.06 $\pm$ 0.17 \\ 
         & RMSE & 70.83 $\pm$ 0.02 & \textbf{68.35 $\pm$ 0.03} & 175.09 $\pm$ 12.25 & 75.17 $\pm$ 0.04 & 75.91 $\pm$ 0.14 & 69.62 $\pm$ 0.23 & 69.28 $\pm$ 0.15 \\ 
         & MAPE(\%) & 39.89 $\pm$ 0.29 & 39.42 $\pm$ 0.10 & 102.80 $\pm$ 22.32 & 44.96 $\pm$ 0.02 & 44.63 $\pm$ 0.37 & 37.82 $\pm$ 0.32 & \textbf{37.33 $\pm$ 0.28} \\ \hline
         \multirow{3}{*}{SXB} & MAE & 78.07 $\pm$ 0.07 & \textbf{75.86 $\pm$ 0.22} & 261.26 $\pm$ 0.00 & 83.83 $\pm$ 0.05 & 83.48 $\pm$ 0.26 & 76.43 $\pm$ 0.07 & 76.10 $\pm$ 0.20 \\ 
         & RMSE & 137.75 $\pm$ 0.13 & \textbf{134.97 $\pm$ 1.08} & 361.96 $\pm$ 0.00 & 146.93 $\pm$ 0.03 & 147.22 $\pm$ 0.28 & 135.84 $\pm$ 0.06 & 136.00 $\pm$ 0.61 \\ 
         & MAPE(\%) & 39.38 $\pm$ 0.52 & 36.88 $\pm$ 0.64 & 223.02 $\pm$ 0.00 & 46.07 $\pm$ 0.29 & 44.95 $\pm$ 0.49 & 38.86 $\pm$ 0.28 & \textbf{36.87 $\pm$ 0.37} \\ \hline
         \multirow{3}{*}{STR} & MAE & 58.31 $\pm$ 0.07 & 55.90 $\pm$ 0.33 & 65.43 $\pm$ 3.82 & 63.52 $\pm$ 0.53 & 67.70 $\pm$ 0.29 & \textbf{55.29 $\pm$ 0.12} & OOM \\ 
         & RMSE & 75.53 $\pm$ 0.02 & 72.34 $\pm$ 0.67 & 86.65 $\pm$ 5.08 & 82.43 $\pm$ 0.47 & 86.71 $\pm$ 0.16 & \textbf{71.79 $\pm$ 0.33} & OOM \\ 
         & MAPE(\%) & 20.16 $\pm$ 0.01 & \textbf{18.61 $\pm$ 0.29} & 21.85 $\pm$ 1.88 & 22.95 $\pm$ 0.11 & 24.03 $\pm$ 0.61 & 18.78 $\pm$ 0.04 & OOM \\ \hline
        \multirow{3}{*}{TPE} & MAE & 134.51 $\pm$ 0.20 & 130.49 $\pm$ 1.84 & 509.35 $\pm$ 13.02 & 138.44 $\pm$ 0.00 & 144.56 $\pm$ 0.25 & \textbf{126.48 $\pm$ 0.29} & 129.85 $\pm$ 0.99 \\ 
         & RMSE & 604.37 $\pm$ 1.42 & 606.52 $\pm$ 6.88 & 1002.98 $\pm$ 12.83 & 592.89 $\pm$ 0.12 & 616.55 $\pm$ 3.87 & \textbf{567.78 $\pm$ 0.27} & 589.03 $\pm$ 1.28 \\ 
         & MAPE(\%) & 45.86 $\pm$ 0.49 & 43.77 $\pm$ 0.72 & 285.05 $\pm$ 4.83 & 44.76 $\pm$ 0.06 & 50.48 $\pm$ 0.57 & \textbf{40.21 $\pm$ 1.83} & 41.77 $\pm$ 1.29 \\ \hline
        \multirow{3}{*}{TO} & MAE & 87.26 $\pm$ 0.28 & \textbf{77.97 $\pm$ 0.72} & 315.29 $\pm$ 0.01 & 80.61 $\pm$ 0.02 & 83.05 $\pm$ 0.09 & 95.23 $\pm$ 1.17 & 97.61 $\pm$ 0.70 \\ 
         & RMSE & 149.60 $\pm$ 0.47 & \textbf{128.32 $\pm$ 1.66} & 421.69 $\pm$ 3.47 & 134.94 $\pm$ 0.10 & 133.55 $\pm$ 0.26 & 150.90 $\pm$ 1.60 & 154.19 $\pm$ 0.30 \\ 
         & MAPE(\%) & 55.62 $\pm$ 0.22 & \textbf{43.26 $\pm$ 0.12} & 390.80 $\pm$ 15.64 & 46.08 $\pm$ 0.69 & 53.08 $\pm$ 1.63 & 56.75 $\pm$ 2.22 & 63.98 $\pm$ 1.64 \\ \hline
        \multirow{3}{*}{YTO} & MAE & 52.35 $\pm$ 0.07 & \textbf{51.04 $\pm$ 0.30} & 148.49 $\pm$ 17.87 & 86.46 $\pm$ 0.04 & 60.73 $\pm$ 0.19 & 57.26 $\pm$ 0.38 & 51.26 $\pm$ 0.14 \\ 
         & RMSE & 88.15 $\pm$ 0.01 & \textbf{83.84 $\pm$ 0.62} & 219.71 $\pm$ 22.22 & 137.24 $\pm$ 0.27 & 95.65 $\pm$ 0.01 & 95.71 $\pm$ 0.74 & 86.95 $\pm$ 0.04 \\ 
         & MAPE(\%) & 38.58 $\pm$ 0.02 & 37.69 $\pm$ 3.84 & 110.54 $\pm$ 44.26 & 65.08 $\pm$ 1.92 & 53.94 $\pm$ 2.03 & 37.79 $\pm$ 0.83 & \textbf{37.62 $\pm$ 1.48} \\ \hline
         \multirow{3}{*}{TLS} & MAE & 257.69 $\pm$ 0.41 & \textbf{255.12 $\pm$ 0.04} & 264.27 $\pm$ 6.31 & 263.95 $\pm$ 0.00 & 294.82 $\pm$ 0.19 & 255.35 $\pm$ 0.01 & 259.64 $\pm$ 0.41 \\ 
         & RMSE & 350.19 $\pm$ 1.23 & 342.33 $\pm$ 0.54 & 352.67 $\pm$ 9.87 & 348.26 $\pm$ 0.09 & 408.19 $\pm$ 0.21 & \textbf{340.68 $\pm$ 0.33} & 351.37 $\pm$ 2.05 \\ 
         & MAPE(\%) & 761.22 $\pm$ 0.81 & 745.90 $\pm$ 7.41 & 792.14 $\pm$ 78.55 & 869.43 $\pm$ 1.90 & 833.62 $\pm$ 1.57 & 749.50 $\pm$ 6.72 & \textbf{730.89 $\pm$ 9.58} \\ \hline
         \multirow{3}{*}{UTC} & MAE & OOM & 66.56 $\pm$ 26.30 & OOM & 48.87 $\pm$ 0.11 & 68.00 $\pm$ 0.85 & 74.93 $\pm$ 0.19 & \textbf{40.09 $\pm$ 0.10} \\          
         & RMSE & OOM & 95.10 $\pm$ 17.24 & OOM & 83.02 $\pm$ 0.25 & 99.18 $\pm$ 0.36 & 122.75 $\pm$ 1.30 & \textbf{72.37 $\pm$ 0.03} \\          
         & MAPE(\%) & OOM & 104.15 $\pm$ 72.84 & OOM & 51.11 $\pm$ 0.40 & 90.66 $\pm$ 5.26 & 89.98 $\pm$ 4.19 & \textbf{37.86 $\pm$ 3.07} \\ \hline         
         \multirow{3}{*}{VNO} & MAE & 87.69 $\pm$ 0.14 & 83.21 $\pm$ 0.40 & OOM & 74.00 $\pm$ 1.45 & 86.63 $\pm$ 0.41 & \textbf{72.98 $\pm$ 0.37} & 96.27 $\pm$ 0.67 \\ 
         & RMSE & 118.99 $\pm$ 0.33 & 113.23 $\pm$ 0.07 & OOM & 100.58 $\pm$ 1.74 & 118.15 $\pm$ 0.43 & \textbf{99.91 $\pm$ 0.11} & 127.62 $\pm$ 0.73 \\ 
         & MAPE(\%) & 53.52 $\pm$ 0.14 & 48.97 $\pm$ 1.46 & OOM & 41.17 $\pm$ 0.65 & 46.50 $\pm$ 0.59 & \textbf{38.67 $\pm$ 1.02} & 64.60 $\pm$ 1.76 \\ \hline
        \multirow{3}{*}{WOB} & MAE & 53.64 $\pm$ 0.29 & 54.06 $\pm$ 3.31 & 58.06 $\pm$ 0.03 & 61.11 $\pm$ 0.00 & 56.56 $\pm$ 0.08 & 53.34 $\pm$ 0.36 & \textbf{52.71 $\pm$ 0.12} \\          
        & RMSE & 84.29 $\pm$ 0.58 & 83.78 $\pm$ 6.06 & 91.34 $\pm$ 0.32 & 95.94 $\pm$ 0.02 & 85.20 $\pm$ 0.03 & 82.52 $\pm$ 0.55 & \textbf{81.89 $\pm$ 0.16} \\          
        & MAPE(\%) & 40.44 $\pm$ 0.19 & \textbf{39.46 $\pm$ 1.23} & 45.09 $\pm$ 0.29 & 49.95 $\pm$ 0.02 & 48.87 $\pm$ 0.60 & 41.26 $\pm$ 0.69 & 40.60 $\pm$ 1.35 \\ \hline
        \multirow{3}{*}{ZRH} & MAE & OOM & 54.55 $\pm$ 0.39 & OOM & 59.12 $\pm$ 0.01 & 58.25 $\pm$ 0.02 & 62.40 $\pm$ 10.21 & \textbf{53.08 $\pm$ 0.23} \\          
        & RMSE & OOM & 77.09 $\pm$ 0.55 & OOM & 83.71 $\pm$ 0.00 & 81.89 $\pm$ 0.11 & 89.30 $\pm$ 15.55 & \textbf{75.22 $\pm$ 0.44} \\          
        & MAPE(\%) & OOM & 35.78 $\pm$ 2.65 & OOM & 43.08 $\pm$ 0.12 & 42.26 $\pm$ 0.44 & 46.00 $\pm$ 11.94 & \textbf{35.18 $\pm$ 0.34} \\ \hline
\bottomrule
\end{longtable}
\end{small}

\begin{small}
\setlength{\tabcolsep}{0.6mm}
\renewcommand{\arraystretch}{1.4}
\begin{longtable}{c|c|c|c|c|c|c|c|c}
\caption{The performance comparison for seven baseline methods over 31 real-world city datasets with \textbf{horizon=12}. The best results in each row are in bold. All experiments are repeated five times, and the mean and standard deviation are reported.}
\label{tab: horizon12}\\
\toprule
     City & Metric & AGCRN & Crossformer & DCRNN & DLinear & FEDformer & GWNet & MTGNN \\ \hline
         \multirow{3}{*}{AGB} & MAE & 55.05 $\pm$ 0.93 & \textbf{49.30 $\pm$ 0.79} & OOM & 56.58 $\pm$ 0.01 & 59.64 $\pm$ 0.20 & 55.81 $\pm$ 1.34 & 53.30 $\pm$ 0.80 \\ 
         & RMSE & 118.56 $\pm$ 0.58 & \textbf{110.08 $\pm$ 0.06} & OOM & 122.58 $\pm$ 0.10 & 132.58 $\pm$ 0.19 & 117.78 $\pm$ 3.74 & 115.28 $\pm$ 0.52 \\ 
         & MAPE(\%) & 51.73 $\pm$ 1.84 & \textbf{41.13 $\pm$ 2.19} & OOM & 44.54 $\pm$ 0.12 & 53.19 $\pm$ 0.31 & 46.17 $\pm$ 2.80 & 44.31 $\pm$ 3.53 \\ \hline
        \multirow{3}{*}{BSL} & MAE & 74.87 $\pm$ 0.54 & 81.11 $\pm$ 3.35 & 120.92 $\pm$ 0.42 & 76.52 $\pm$ 0.22 & \textbf{66.00 $\pm$ 0.70} & 97.61 $\pm$ 4.34 & 89.52 $\pm$ 0.36 \\ 
         & RMSE & 110.37 $\pm$ 0.82 & 122.17 $\pm$ 4.15 & 160.46 $\pm$ 2.57 & 126.12 $\pm$ 0.48 & \textbf{96.52 $\pm$ 0.35} & 139.03 $\pm$ 8.66 & 126.94 $\pm$ 0.56 \\ 
         & MAPE(\%) & 65.54 $\pm$ 0.65 & 89.04 $\pm$ 5.12 & 183.70 $\pm$ 15.83 & \textbf{62.84 $\pm$ 0.30} & 69.59 $\pm$ 3.62 & 129.70 $\pm$ 6.69 & 111.43 $\pm$ 0.24 \\ \hline
        \multirow{3}{*}{BRN} & MAE & \textbf{53.50 $\pm$ 2.10} & 54.20 $\pm$ 1.85 & OOM & 56.13 $\pm$ 0.01 & 61.14 $\pm$ 0.33 & 54.13 $\pm$ 0.47 & 84.36 $\pm$ 1.26 \\ 
         & RMSE & 500.59 $\pm$ 21.34 & 483.08 $\pm$ 0.25 & OOM & 495.21 $\pm$ 0.17 & 489.57 $\pm$ 0.68 & \textbf{479.82 $\pm$ 1.60} & 553.53 $\pm$ 12.87 \\ 
         & MAPE(\%) & 245.52 $\pm$ 4.25 & \textbf{202.93 $\pm$ 62.89} & OOM & 290.64 $\pm$ 1.52 & 307.33 $\pm$ 16.91 & 340.44 $\pm$ 23.69 & 513.55 $\pm$ 31.06 \\ \hline
        \multirow{3}{*}{BHX} & MAE & 146.03 $\pm$ 0.35 & 102.03 $\pm$ 15.50 & 282.12 $\pm$ 13.73 & 131.89 $\pm$ 9.65 & 140.17 $\pm$ 0.70 & 129.19 $\pm$ 2.55 & \textbf{101.51 $\pm$ 0.10} \\          
        & RMSE & 208.28 $\pm$ 0.65 & 155.04 $\pm$ 22.46 & 384.96 $\pm$ 40.10 & 193.75 $\pm$ 6.56 & 214.60 $\pm$ 0.16 & 192.02 $\pm$ 0.76 & \textbf{154.51 $\pm$ 0.86} \\          
        & MAPE(\%) & 109.60 $\pm$ 0.28 & 63.90 $\pm$ 4.44 & 279.89 $\pm$ 61.60 & 90.05 $\pm$ 7.98 & 86.17 $\pm$ 0.27 & 80.07 $\pm$ 9.65 & \textbf{58.53 $\pm$ 2.38} \\ \hline        
        \multirow{3}{*}{BOL} & MAE & \textbf{33.38 $\pm$ 0.12} & 35.32 $\pm$ 0.68 & 45.53 $\pm$ 0.92 & 42.42 $\pm$ 0.09 & 44.03 $\pm$ 0.11 & 39.58 $\pm$ 2.22 & 34.94 $\pm$ 0.49 \\ 
         & RMSE & 93.57 $\pm$ 0.67 & 93.44 $\pm$ 2.09 & 106.98 $\pm$ 0.01 & 102.71 $\pm$ 0.26 & 105.61 $\pm$ 0.31 & 97.56 $\pm$ 0.15 & \textbf{92.74 $\pm$ 0.59}\\
         & MAPE(\%) & 23.09 $\pm$ 0.63 & \textbf{22.13 $\pm$ 0.46} & 33.86 $\pm$ 1.51 & 38.07 $\pm$ 0.13 & 33.57 $\pm$ 0.61 & 27.38 $\pm$ 1.10 & 22.27 $\pm$ 0.29 \\ \hline
        \multirow{3}{*}{BOD} & MAE & 84.17 $\pm$ 0.76 & \textbf{78.48 $\pm$ 2.18} & 234.94 $\pm$ 3.51 & 78.76 $\pm$ 0.01 & 87.29 $\pm$ 0.40 & 91.20 $\pm$ 1.25 & 109.35 $\pm$ 1.40 \\ 
         & RMSE & 138.13 $\pm$ 1.65 & 125.87 $\pm$ 0.23 & 312.43 $\pm$ 6.74 & \textbf{122.38 $\pm$ 0.08} & 131.87 $\pm$ 0.90 & 139.59 $\pm$ 1.34 & 173.27 $\pm$ 0.92 \\ 
         & MAPE(\%) & 44.29 $\pm$ 0.82 & \textbf{41.19 $\pm$ 1.49} & 278.32 $\pm$ 1.47 & 54.23 $\pm$ 0.18 & 59.39 $\pm$ 0.18 & 73.00 $\pm$ 2.49 & 74.25 $\pm$ 5.28 \\ \hline
        \multirow{3}{*}{BRE} & MAE & \textbf{57.33 $\pm$ 0.09} & 59.38 $\pm$ 0.08 & OOM & 68.48 $\pm$ 0.01 & 64.98 $\pm$ 0.09 & 58.62 $\pm$ 0.06 & 58.05 $\pm$ 0.39 \\ 
         & RMSE & \textbf{95.67 $\pm$ 0.22} & 98.02 $\pm$ 0.06 & OOM & 111.31 $\pm$ 0.02 & 103.81 $\pm$ 0.10 & 96.37 $\pm$ 0.06 & 96.10 $\pm$ 0.84 \\ 
         & MAPE(\%) & 36.89 $\pm$ 0.01 & \textbf{35.19 $\pm$ 0.34} & OOM & 45.77 $\pm$ 0.22 & 44.95 $\pm$ 0.13 & 38.23 $\pm$ 0.29 & 36.70 $\pm$ 0.29 \\ \hline
        \multirow{3}{*}{KN} & MAE & OOM & 52.98 $\pm$ 1.36 & 116.61 $\pm$ 0.23 & \textbf{42.78 $\pm$ 0.05} & 44.75 $\pm$ 0.24 & 49.44 $\pm$ 0.46 & 57.22 $\pm$ 1.96 \\ 
         & RMSE & OOM & 83.22 $\pm$ 0.84 & 149.74 $\pm$ 0.66 & 67.32 $\pm$ 0.14 & \textbf{65.83 $\pm$ 0.35} & 79.03 $\pm$ 0.50 & 86.98 $\pm$ 2.16 \\ 
         & MAPE(\%) & OOM & \textbf{54.05 $\pm$ 3.83} & 297.19 $\pm$ 1.49 & 70.57 $\pm$ 0.27 & 77.29 $\pm$ 3.06 & 61.92 $\pm$ 1.39 & 89.88 $\pm$ 6.97 \\ \hline
        \multirow{3}{*}{DA} & MAE & 56.88 $\pm$ 0.01 & \textbf{56.19 $\pm$ 0.15} & OOM & 59.55 $\pm$ 0.01 & 61.34 $\pm$ 0.24 & 57.52 $\pm$ 0.40 & 57.85 $\pm$ 0.09 \\ 
         & RMSE & 88.16 $\pm$ 0.02 & \textbf{86.45 $\pm$ 0.70} & OOM & 90.81 $\pm$ 0.06 & 90.98 $\pm$ 0.34 & 87.83 $\pm$ 1.01 & 90.13 $\pm$ 0.16 \\ 
         & MAPE(\%) & 53.29 $\pm$ 0.09 & 54.39 $\pm$ 6.00 & OOM & 57.20 $\pm$ 0.02 & 66.56 $\pm$ 0.19 & 55.03 $\pm$ 0.35 & \textbf{52.07 $\pm$ 0.04} \\ \hline
         \multirow{3}{*}{ESS} & MAE & 44.08 $\pm$ 0.27 & 44.91 $\pm$ 0.32 & 179.03 $\pm$ 1.10 & 62.74 $\pm$ 0.04 & 49.97 $\pm$ 0.14 & 41.68 $\pm$ 0.05 & \textbf{41.42 $\pm$ 0.27}\\          
         & RMSE & 65.43 $\pm$ 0.68 & 67.98 $\pm$ 0.93 & 232.52 $\pm$ 0.26 & 92.85 $\pm$ 0.16 & 73.19 $\pm$ 0.40 & \textbf{62.40 $\pm$ 0.24} & 63.04 $\pm$ 0.50 \\ 
         & MAPE(\%) & \textbf{35.95 $\pm$ 1.15} & 49.74 $\pm$ 5.81 & 308.94 $\pm$ 8.39 & 55.33 $\pm$ 1.11 & 48.59 $\pm$ 1.21 & 37.57 $\pm$ 0.39 & 38.41 $\pm$ 2.97 \\ \hline
        \multirow{3}{*}{FRA} & MAE & 205.72 $\pm$ 0.83 & 205.31 $\pm$ 22.15 & 168.51 $\pm$ 1.59 & \textbf{109.03 $\pm$ 36.74} & 132.32 $\pm$ 1.38 & 250.59 $\pm$ 11.89 & 386.51 $\pm$ 26.96 \\ 
         & RMSE & 238.54 $\pm$ 0.79 & 234.38 $\pm$ 24.27 & 209.58 $\pm$ 0.03 & \textbf{132.51 $\pm$ 38.23} & 165.66 $\pm$ 0.15 & 282.51 $\pm$ 11.51 & 408.51 $\pm$ 27.19 \\ 
         & MAPE(\%) & 73.26 $\pm$ 0.25 & 70.54 $\pm$ 5.11 & 55.50 $\pm$ 0.97 & \textbf{35.27 $\pm$ 11.29} & 41.22 $\pm$ 0.39 & 85.66 $\pm$ 5.06 & 132.85 $\pm$ 10.28 \\ \hline
        \multirow{3}{*}{GRZ} & MAE & 63.46 $\pm$ 0.17 & \textbf{57.03 $\pm$ 0.76} & 191.54 $\pm$ 0.00 & 71.93 $\pm$ 0.01 & 63.14 $\pm$ 0.03 & 62.37 $\pm$ 0.78 & 59.51 $\pm$ 0.28 \\ 
         & RMSE & 95.67 $\pm$ 0.07 & \textbf{83.99 $\pm$ 0.61} & 241.77 $\pm$ 0.00 & 103.75 $\pm$ 0.01 & 91.38 $\pm$ 0.00 & 92.58 $\pm$ 0.82 & 95.09 $\pm$ 0.61 \\ 
         & MAPE(\%) & 109.30 $\pm$ 3.51 & 77.93 $\pm$ 5.43 & 460.74 $\pm$ 0.00 & 81.48 $\pm$ 0.13 & 78.44 $\pm$ 0.27 & \textbf{70.16 $\pm$ 4.58} & 77.40 $\pm$ 2.29 \\ \hline
        \multirow{3}{*}{GRQ} & MAE & 77.05 $\pm$ 0.27 & \textbf{72.08 $\pm$ 3.04} & 150.91 $\pm$ 0.00 & 72.34 $\pm$ 1.76 & 83.37 $\pm$ 2.14 & 72.33 $\pm$ 0.77 & 82.93 $\pm$ 6.04 \\ 
         & RMSE & 104.16 $\pm$ 0.05 & \textbf{98.38 $\pm$ 2.63} & 205.49 $\pm$ 0.00 & 100.03 $\pm$ 0.44 & 115.72 $\pm$ 2.69 & 100.25 $\pm$ 0.83 & 110.28 $\pm$ 7.48 \\ 
         & MAPE(\%) & 41.63 $\pm$ 0.69 & \textbf{35.60 $\pm$ 1.85} & 117.24 $\pm$ 0.00 & 44.18 $\pm$ 2.38 & 46.94 $\pm$ 1.33 & 38.69 $\pm$ 0.88 & 45.80 $\pm$ 2.96 \\ \hline
        \multirow{3}{*}{HAM} & MAE & 47.08 $\pm$ 0.23 & 45.88 $\pm$ 0.41 & 97.51 $\pm$ 0.59 & 48.97 $\pm$ 0.02 & 49.77 $\pm$ 0.01 & \textbf{45.35 $\pm$ 0.05} & 46.06 $\pm$ 0.01 \\ 
         & RMSE & 80.67 $\pm$ 0.53 & 78.49 $\pm$ 1.08 & 150.82 $\pm$ 2.57 & 84.20 $\pm$ 0.08 & 83.90 $\pm$ 0.06 & \textbf{77.70 $\pm$ 0.01} & 82.77 $\pm$ 2.08 \\ 
         & MAPE(\%) & 45.78 $\pm$ 0.83 & \textbf{44.31 $\pm$ 3.07} & 111.41 $\pm$ 4.58 & 52.32 $\pm$ 0.20 & 53.06 $\pm$ 0.21 & 44.81 $\pm$ 0.31 & 45.29 $\pm$ 0.06 \\ \hline
        \multirow{3}{*}{INN} & MAE & 75.51 $\pm$ 0.80 & 72.05 $\pm$ 0.00 & 347.11 $\pm$ 2.98 & 106.94 $\pm$ 0.00 & 80.50 $\pm$ 1.00 & \textbf{69.08 $\pm$ 0.43} & OOM \\ 
         & RMSE & 110.87 $\pm$ 0.65 & 106.99 $\pm$ 0.06 & 464.48 $\pm$ 8.39 & 173.49 $\pm$ 0.15 & 117.77 $\pm$ 1.61 & \textbf{100.59 $\pm$ 0.79} & OOM \\ 
         & MAPE(\%) & 31.27 $\pm$ 1.22 & 36.38 $\pm$ 0.72 & 300.48 $\pm$ 37.49 & 46.75 $\pm$ 0.75 & 34.23 $\pm$ 0.15 & \textbf{27.74 $\pm$ 0.06} & OOM \\ \hline
        \multirow{3}{*}{KS} & MAE & 97.97 $\pm$ 0.72 & 103.19 $\pm$ 5.03 & 227.96 $\pm$ 0.00 & \textbf{78.86 $\pm$ 14.60} & 104.59 $\pm$ 1.19 & 80.04 $\pm$ 0.72 & 224.87 $\pm$ 1.62 \\ 
         & RMSE & 244.74 $\pm$ 0.48 & 244.15 $\pm$ 6.33 & 332.77 $\pm$ 0.00 & \textbf{188.64 $\pm$ 2.13} & 209.20 $\pm$ 1.15 & 195.69 $\pm$ 0.26 & 314.80 $\pm$ 4.13 \\ 
         & MAPE(\%) & 143.62 $\pm$ 1.84 & 156.53 $\pm$ 13.18 & 433.64 $\pm$ 0.00 & 110.39 $\pm$ 41.23 & 162.29 $\pm$ 2.84 & \textbf{109.72 $\pm$ 2.55} & 407.04 $\pm$ 1.53 \\ \hline
        \multirow{3}{*}{MAN} & MAE & 117.20 $\pm$ 2.55 & 109.56 $\pm$ 2.55 & 336.84 $\pm$ 0.00 & \textbf{103.86 $\pm$ 9.84} & 123.20 $\pm$ 0.27 & 107.88 $\pm$ 1.14 & 107.97 $\pm$ 2.60 \\ 
         & RMSE & 196.04 $\pm$ 4.11 & 190.75 $\pm$ 4.72 & 450.82 $\pm$ 0.00 & \textbf{173.23 $\pm$ 12.91} & 198.08 $\pm$ 0.51 & 181.98 $\pm$ 4.00 & 186.65 $\pm$ 3.39 \\ 
         & MAPE(\%) & 47.38 $\pm$ 0.86 & 45.64 $\pm$ 2.89 & 278.97 $\pm$ 0.00 & 46.64 $\pm$ 3.55 & 60.39 $\pm$ 0.14 & \textbf{44.84 $\pm$ 1.22} & 47.02 $\pm$ 1.75 \\ \hline
        \multirow{3}{*}{MEL} & MAE & 64.79 $\pm$ 0.03 & \textbf{54.63 $\pm$ 4.48} & OOM & 87.95 $\pm$ 1.60 & 67.32 $\pm$ 0.31 & 69.54 $\pm$ 1.08 & 56.73 $\pm$ 0.94 \\ 
         & RMSE & 101.09 $\pm$ 0.06 & \textbf{81.23 $\pm$ 6.52} & OOM & 127.60 $\pm$ 1.41 & 97.39 $\pm$ 0.04 & 105.83 $\pm$ 1.53 & 87.59 $\pm$ 0.91 \\ 
         & MAPE(\%) & 55.34 $\pm$ 0.02 & 49.10 $\pm$ 5.14 & OOM & 103.87 $\pm$ 4.79 & 70.74 $\pm$ 1.14 & \textbf{47.17 $\pm$ 0.20} & 50.21 $\pm$ 0.12 \\ \hline
        \multirow{3}{*}{RTM} & MAE & \textbf{57.11 $\pm$ 0.15} & 58.08 $\pm$ 1.24 & 189.90 $\pm$ 0.00 & 85.30 $\pm$ 0.50 & 82.23 $\pm$ 0.36 & 85.20 $\pm$ 1.78 & 65.01 $\pm$ 0.31 \\ 
         & RMSE & 99.90 $\pm$ 0.34 & \textbf{98.98 $\pm$ 0.53} & 250.70 $\pm$ 0.00 & 141.23 $\pm$ 0.46 & 128.17 $\pm$ 0.09 & 136.93 $\pm$ 2.26 & 110.64 $\pm$ 0.98 \\ 
         & MAPE(\%) & \textbf{43.47 $\pm$ 0.63} & 55.83 $\pm$ 11.21 & 373.03 $\pm$ 0.00 & 65.00 $\pm$ 0.87 & 76.28 $\pm$ 0.17 & 62.30 $\pm$ 2.15 & 46.88 $\pm$ 1.28 \\ \hline
        \multirow{3}{*}{SDR} & MAE & 121.60 $\pm$ 0.24 & 129.77 $\pm$ 6.53 & 272.16 $\pm$ 0.00 & 120.00 $\pm$ 0.03 & 152.27 $\pm$ 0.77 & \textbf{104.78 $\pm$ 0.34} & 119.57 $\pm$ 0.47 \\ 
         & RMSE & 278.70 $\pm$ 0.41 & 279.99 $\pm$ 1.64 & 440.08 $\pm$ 0.00 & 262.91 $\pm$ 0.01 & 284.60 $\pm$ 0.05 & \textbf{249.10 $\pm$ 0.76} & 274.87 $\pm$ 3.75 \\ 
         & MAPE(\%) & 67.07 $\pm$ 2.54 & 92.03 $\pm$ 14.14 & 301.72 $\pm$ 0.00 & 68.81 $\pm$ 0.28 & 119.83 $\pm$ 1.98 & 58.89 $\pm$ 5.49 & \textbf{55.51 $\pm$ 0.36} \\ \hline
         \multirow{3}{*}{SP} & MAE & 49.26 $\pm$ 0.11 & \textbf{48.39 $\pm$ 0.19} & 122.84 $\pm$ 4.61 & 55.70 $\pm$ 0.01 & 54.59 $\pm$ 0.03 & 49.12 $\pm$ 0.17 & 48.70 $\pm$ 0.25 \\ 
         & RMSE & 70.97 $\pm$ 0.13 & \textbf{69.48 $\pm$ 0.62} & 175.83 $\pm$ 13.31 & 79.94 $\pm$ 0.01 & 77.28 $\pm$ 0.03 & 70.95 $\pm$ 0.34 & 70.28 $\pm$ 0.31 \\ 
         & MAPE(\%) & 39.90 $\pm$ 0.46 & 38.25 $\pm$ 1.95 & 102.22 $\pm$ 24.13 & 47.80 $\pm$ 0.10 & 46.68 $\pm$ 0.32 & 38.51 $\pm$ 0.41 & \textbf{37.52 $\pm$ 0.37} \\ \hline
         \multirow{3}{*}{SXB} & MAE & 80.33 $\pm$ 0.24 & \textbf{78.32 $\pm$ 0.23} & 262.40 $\pm$ 0.00 & 94.09 $\pm$ 0.05 & 89.88 $\pm$ 0.17 & 80.19 $\pm$ 0.36 & 78.71 $\pm$ 0.29 \\ 
         & RMSE & 142.42 $\pm$ 0.73 & \textbf{139.91 $\pm$ 0.95} & 363.49 $\pm$ 0.00 & 162.33 $\pm$ 0.05 & 156.75 $\pm$ 0.32 & 141.91 $\pm$ 0.63 & 141.32 $\pm$ 0.36 \\ 
         & MAPE(\%) & 40.29 $\pm$ 0.54 & 38.22 $\pm$ 0.97 & 223.27 $\pm$ 0.00 & 51.18 $\pm$ 0.37 & 49.09 $\pm$ 0.17 & 40.96 $\pm$ 0.76 & \textbf{37.89 $\pm$ 0.11} \\ \hline
         \multirow{3}{*}{STR} & MAE & 61.10 $\pm$ 0.38 & 57.78 $\pm$ 0.84 & 71.86 $\pm$ 5.01 & 73.59 $\pm$ 0.94 & 68.30 $\pm$ 0.33 & \textbf{56.40 $\pm$ 0.08} & OOM \\ 
         & RMSE & 81.05 $\pm$ 0.55 & 74.94 $\pm$ 1.18 & 96.17 $\pm$ 7.97 & 96.18 $\pm$ 1.48 & 90.50 $\pm$ 0.70 & \textbf{73.46 $\pm$ 0.06} & OOM \\ 
         & MAPE(\%) & 22.32 $\pm$ 0.19 & 21.28 $\pm$ 1.87 & 25.56 $\pm$ 2.23 & 30.23 $\pm$ 0.66 & 25.02 $\pm$ 1.17 & \textbf{20.36 $\pm$ 0.09} & OOM \\ \hline
        \multirow{3}{*}{TPE} & MAE & 147.54 $\pm$ 0.61 & \textbf{140.98 $\pm$ 2.89} & 499.95 $\pm$ 8.83 & 163.96 $\pm$ 0.02 & 168.31 $\pm$ 0.17 & 142.71 $\pm$ 0.55 & 142.86 $\pm$ 1.83 \\ 
         & RMSE & 670.14 $\pm$ 0.45 & 671.29 $\pm$ 2.58 & 996.04 $\pm$ 7.32 & 726.80 $\pm$ 0.40 & 726.73 $\pm$ 1.06 & \textbf{666.42 $\pm$ 0.36} & 678.02 $\pm$ 3.41 \\ 
         & MAPE(\%) & 54.90 $\pm$ 0.81 & 47.50 $\pm$ 4.82 & 280.63 $\pm$ 13.88 & 52.22 $\pm$ 0.17 & 60.40 $\pm$ 0.77 & 45.94 $\pm$ 2.11 & \textbf{44.89 $\pm$ 0.61} \\ \hline
         \multirow{3}{*}{TO} & MAE & 104.21 $\pm$ 0.69 & \textbf{92.97 $\pm$ 2.72} & 319.07 $\pm$ 5.97 & 106.30 $\pm$ 0.02 & 105.57 $\pm$ 0.66 & 139.93 $\pm$ 0.61 & 133.21 $\pm$ 1.02 \\ 
         & RMSE & 178.09 $\pm$ 1.38 & \textbf{154.59 $\pm$ 2.71} & 424.02 $\pm$ 8.74 & 182.49 $\pm$ 0.32 & 167.92 $\pm$ 0.92 & 231.65 $\pm$ 2.32 & 221.49 $\pm$ 0.33 \\ 
         & MAPE(\%) & 67.26 $\pm$ 0.52 & \textbf{51.82 $\pm$ 1.26} & 400.37 $\pm$ 6.15 & 59.46 $\pm$ 1.38 & 69.03 $\pm$ 0.16 & 82.81 $\pm$ 0.41 & 87.58 $\pm$ 0.12 \\ \hline
        \multirow{3}{*}{YTO} & MAE & 57.16 $\pm$ 0.24 & \textbf{56.66 $\pm$ 1.67} & 173.86 $\pm$ 9.18 & 125.57 $\pm$ 0.08 & 74.96 $\pm$ 1.04 & 70.85 $\pm$ 0.31 & 60.80 $\pm$ 0.21 \\ 
         & RMSE & 94.09 $\pm$ 0.11 & \textbf{90.75 $\pm$ 1.41} & 253.42 $\pm$ 9.38 & 191.73 $\pm$ 0.25 & 114.34 $\pm$ 1.97 & 115.62 $\pm$ 0.60 & 100.24 $\pm$ 0.45 \\ 
         & MAPE(\%) & 44.91 $\pm$ 1.53 & \textbf{43.39 $\pm$ 4.12} & 128.97 $\pm$ 45.99 & 110.73 $\pm$ 1.97 & 73.18 $\pm$ 1.56 & 44.39 $\pm$ 0.43 & 43.66 $\pm$ 0.17 \\ \hline
        \multirow{3}{*}{TLS} & MAE & 257.57 $\pm$ 0.47 & 255.17 $\pm$ 0.18 & 264.44 $\pm$ 6.13 & 264.06 $\pm$ 0.00 & 298.73 $\pm$ 0.16 & \textbf{255.11 $\pm$ 0.01} & 257.76 $\pm$ 0.30 \\ 
         & RMSE & 348.95 $\pm$ 1.83 & 342.20 $\pm$ 1.02 & 352.99 $\pm$ 9.76 & 348.31 $\pm$ 0.02 & 408.17 $\pm$ 0.23 & \textbf{340.28 $\pm$ 0.14} & 346.83 $\pm$ 0.97 \\ 
         & MAPE(\%) & 746.15 $\pm$ 8.05 & 758.84 $\pm$ 36.31 & 791.75 $\pm$ 78.81 & 867.12 $\pm$ 0.77 & 842.35 $\pm$ 8.43 & \textbf{745.07 $\pm$ 5.77} & 746.42 $\pm$ 0.82 \\ \hline
        \multirow{3}{*}{UTC} & MAE & OOM & 55.13 $\pm$ 4.47 & OOM & 57.76 $\pm$ 0.72 & 73.45 $\pm$ 1.74 & 73.81 $\pm$ 0.57 & \textbf{40.36 $\pm$ 0.11}\\          
        & RMSE & OOM & 89.55 $\pm$ 1.67 & OOM & 97.52 $\pm$ 1.24 & 109.98 $\pm$ 1.82 & 123.95 $\pm$ 2.23 & \textbf{75.86 $\pm$ 0.31}\\          
        & MAPE(\%) & OOM & 80.70 $\pm$ 18.35 & OOM & 64.93 $\pm$ 0.77 & 105.19 $\pm$ 2.55 & 93.66 $\pm$ 5.16 & \textbf{40.85 $\pm$ 2.82} \\ \hline         
        \multirow{3}{*}{VNO} & MAE & 96.16 $\pm$ 0.15 & 91.31 $\pm$ 0.19 & OOM & 81.71 $\pm$ 4.58 & 98.47 $\pm$ 0.10 & \textbf{80.01 $\pm$ 0.59} & 105.63 $\pm$ 1.55 \\ 
         & RMSE & 130.79 $\pm$ 0.34 & 124.11 $\pm$ 0.44 & OOM & 111.45 $\pm$ 2.76 & 131.28 $\pm$ 0.02 & \textbf{109.93 $\pm$ 0.18} & 141.02 $\pm$ 1.91 \\ 
         & MAPE(\%) & 63.46 $\pm$ 0.21 & 58.74 $\pm$ 1.31 & OOM & 47.91 $\pm$ 8.96 & 60.49 $\pm$ 0.08 & \textbf{46.17 $\pm$ 1.50} & 77.00 $\pm$ 2.31 \\ \hline
        \multirow{3}{*}{WOB} & MAE & 56.73 $\pm$ 0.26 & \textbf{53.73 $\pm$ 0.93} & 65.57 $\pm$ 0.76 & 69.50 $\pm$ 0.05 & 61.35 $\pm$ 0.11 & 57.82 $\pm$ 0.18 & 56.30 $\pm$ 0.43 \\ 
         & RMSE & 90.97 $\pm$ 0.48 & \textbf{83.48 $\pm$ 2.09} & 104.61 $\pm$ 1.31 & 111.82 $\pm$ 0.01 & 92.81 $\pm$ 0.13 & 91.64 $\pm$ 0.05 & 89.31 $\pm$ 0.69 \\ 
         & MAPE(\%) & 42.48 $\pm$ 1.04 & \textbf{41.16 $\pm$ 3.01} & 53.82 $\pm$ 1.86 & 55.70 $\pm$ 0.59 & 54.53 $\pm$ 0.12 & 44.75 $\pm$ 0.61 & 43.14 $\pm$ 1.57 \\ \hline
        \multirow{3}{*}{ZRH} & MAE & OOM & 56.73 $\pm$ 0.03 & OOM & 65.62 $\pm$ 0.00 & 65.29 $\pm$ 0.20 & 61.77 $\pm$ 4.83 & \textbf{54.71 $\pm$ 0.20}\\          
        & RMSE & OOM & 80.80 $\pm$ 0.23 & OOM & 94.06 $\pm$ 0.01 & 91.54 $\pm$ 0.20 & 88.03 $\pm$ 6.88 & \textbf{77.75 $\pm$ 0.44}\\          
        & MAPE(\%) & OOM & 36.22 $\pm$ 0.43 & OOM & 47.73 $\pm$ 0.13 & 48.20 $\pm$ 0.21 & 46.66 $\pm$ 8.57 & \textbf{35.87 $\pm$ 0.06} \\ \hline 
\bottomrule
\end{longtable}
\end{small}

\section{Limitations}
\begin{itemize}
    \item \textbf{Crowdsourced OSM quality and geographic bias.} OSM+ inherits the strengths and limitations of OpenStreetMap. Coverage and tag consistency vary across regions because OSM is edited by a distributed volunteer community. Regions with rich OSM activity often have dense road networks and detailed attributes, while other regions may miss minor roads or contain generic labels. This bias can affect both direct graph analysis and downstream model evaluation. We therefore recommend reporting coverage statistics and, when possible, stratifying evaluation by data-quality level.

    \item \textbf{Snapshot freshness.} The experiments in this paper use a fixed March 2026 snapshot for reproducibility. As OSM evolves, this snapshot will become outdated for operational applications. Our monthly versioning plan and released preprocessing pipeline mitigate this issue, but users should select the version that matches their intended temporal scope.

    \item \textbf{Traffic benchmark scope.} The 31-city traffic-prediction benchmark is constructed from UTD19 historical observations aligned to OSM+. It is an offline research benchmark rather than an online traffic-management ground truth. Source-specific issues in UTD19, such as missing measurements, latency, or sensor-quality variation, are inherited by the benchmark.

    \item \textbf{Sensor-to-road matching noise.} The nearest-edge alignment between traffic observations and OSM+ road segments can introduce noise, especially in regions where minor roads are missing or OSM coverage is sparse. This noise is realistic for cross-city geospatial benchmarking, but it should not be ignored when interpreting model failures or unusually high errors.

    \item \textbf{External covariates.} Weather, events, POIs, and other contextual variables can improve traffic forecasting. The current benchmark intentionally uses historical traffic observations and road-network structure to keep the 31-city setting consistent and reproducible. OSM+ supports joining with such external covariates, and future benchmark versions can extend the task setup accordingly.

    \item \textbf{Application-specific APIs.} Although we release basic APIs for querying and processing OSM+, individual applications may require additional preprocessing or domain-specific validation. We encourage researchers to contribute additional converters, audits, and downstream benchmark tasks.
\end{itemize}

\section{Author Statement}
We want to show our great thanks to OpenStreetMap~\citep{haklay2008openstreetmap}, UTD19~\citep{loder2020utd19}, Oak Ridge National Laboratory~\citep{Sims2023-zc} and NASA Earth Observatory~\citep{weier2003nasa} for providing free-public roadnet data, traffic flow data, population data and nightlight data.  Please let us know if any issues are found. We will take appropriate action when needed, e.g, to remove data records with such issues.


\end{document}